\documentclass[aps,twocolumn]{revtex4}

\usepackage{mathrsfs}
\usepackage{amsmath}
\usepackage{graphicx}
\usepackage{epstopdf}

\begin{document}

\title{A sticky business: the status of the conjectured viscosity/entropy density bound}

\author{Aleksey Cherman}
\email{alekseyc@physics.umd.com} \affiliation{Department of Physics,
University of Maryland, College Park, MD 20742-4111}

\author{Thomas D. Cohen}
\email{cohen@physics.umd.edu}

\affiliation{Department of Physics, University of Maryland,
College Park, MD 20742-4111}

\author{Paul M. Hohler}
\email{pmhohler@physics.umd.edu}

\affiliation{Department of Physics, University of Maryland,
College Park, MD 20742-4111}

\begin{abstract}
There have been a number of forms of a conjecture that there is a
universal lower bound on the ratio, $\eta/s$, of the shear
viscosity, $\eta$, to entropy density, $s$, with several different
domains of validity. We examine the various forms of the
conjecture.  We argue that a number of variants of the conjecture
are not viable due to the existence of theoretically consistent
counterexamples. We also note that much of the evidence in favor
of a bound does not apply to the variants which have not yet been
ruled out.

\end{abstract}
\maketitle

\section{Introduction}
\label{sect:intro}
Kovtun, Son, and Starinets (KSS) have proposed a conjecture that there
is a universal bound for the ratio of shear viscosity, $\eta$, to
entropy density, $s$, \cite{KSS1}:
\begin{equation} \label{eq:bound}
\frac{\eta}{s} \geq \frac{\hbar}{k_B}\frac{1}{4 \pi },
\end{equation}
where $\hbar$ and $k_B$ are Plank's constant and Boltzmann's
constant, respectively.  (For the remainder of this paper we will
use units with $\hbar=1$ and $k_B=1$.) KSS found that
Eq.~(\ref{eq:bound}) is saturated by certain strongly coupled
field theories which have a super-gravity dual \cite{KSS1}, and
conjectured that $\eta/s$ has a universal lower limit. Physically
interesting and accessible fluids, such as water, liquid nitrogen,
and helium-4 satisfy the bound \cite{KSS2}. The bound appears to
be well justified for the class of field theories originally
considered by KSS \cite{KSS1}, but it is not obvious from first
principles that it should apply more universally (hence its status
as a conjecture).

The original form of the KSS conjecture states that the bound
should be universal and apply to \emph{all} fluids, including
non-relativistic fluids \cite{KSS1}. Yet even such an
all-encompassing statement includes ambiguities. It is not clear
what one might mean by ``all fluids" in such a context. Is the
conjecture limited to physically realizable systems, or is it
equally applicable to theoretical fluids which can be constructed
in some given class of theories? If so, in which class of theories
does the bound hold? Is a ``fluid" required to be absolutely
stable, or can the fluid be metastable? Are the number of species
of particle that compose the fluid limited? Perhaps due to
questions such as these, a number of variants of the conjecture
with various proposed domains of validity were subsequently
proposed by KSS. These include variants which stipulate that the
bound is valid for ``all relativistic quantum field theories at
finite temperature and zero chemical potential'' \cite{KSS2}, for
at least a ``single component nonrelativistic gas of particles
with either spin zero or spin 1/2'' \cite{KSS2}, or for ``all
systems which can be obtained from a sensible relativistic quantum
field theory by turning on temperatures and chemical potentials''
\cite{SS1}. While some of these variants appear quite similar at
first glance, they actually have quite different regimes of
validity.

If the bound could be shown to be correct in any of its proposed
forms, or indeed in some readily specifiable alternative form, it
would represent a truly major advance in our understanding of
quantum many-body physics. Indeed, even as a conjecture it has
been invoked in discussing systems as diverse as ultra-cold gases
of trapped atoms \cite{atom} and the quark-gluon plasma (QGP)
\cite{qgp}. Since KSS first conjectured their bound, the ratio of
shear viscosity to entropy density has been investigated in a
variety of systems,
\cite{laha,dobado,chen1,shafer,liu,gj,chen2,cohen,spain}. The
smallest reported measurement of $\eta/s$ has been associated with
the QGP at RHIC \cite{qgp}. (A more recent analysis of the
data from RHIC may actually be consistent with a violation of the
proposed bound \cite{rr}.)  Since the $\eta/s$ bound may (or may
not) have a rather extensive scope, it is important to understand
 in which types of systems one should expect the bound to
hold.

As will be discussed in some detail below, the conjectured domains
of validity of the conjecture differ radically from form to form.
Moreover, apparently innocuous changes in the formulation of the
variants of the conjecture can radically alter the systems for
which they apply. Accordingly, it is important in dealing with
this subject to clarify the precise nature of the various forms of
the conjecture and, in particular, to which physical systems they
might apply.

The outline of this paper is as follows. In
Sec.~\ref{sect:evidence}, we begin with a brief discussion of
evidence in favor of the KSS bound in any of its forms.  In
Sec.~\ref{sect:classification}, we classify a set of possible
domains of applicability for which Eq.~(\ref{eq:bound}) might
hold. The various forms of the conjecture proposed by KSS will
form a subset of these. Having delineated the various forms, we
critically examine the physical systems for which these variants
actually apply. In Secs.~\ref{sect:class1}, \ref{sect:class2}, and
\ref{sect:class3}, we address the key issue of the evidence that
any particular variation of the conjecture might be valid. A
natural question in this context is whether one can construct a
theoretical counterexample to a particular variant. In these
sections, we will present counterexamples to a number of variants
of the conjecture.  In this context, we discuss in Sec.~VIC a
subtle issue raised in Ref.~\cite{sonComment} regarding the
interplay of thermodynamic and hydrodynamic limits for the
counter-example in Sec.~\ref{sect:class3} (a heavy meson system
based on a UV-complete quantum field theory).  As will be seen,
while Ref.~\cite{sonComment} raises a profound issue, ultimately,
it does not invalidate the counterexample.

In these sections we also point out that much of the evidence
which seems to support the conjecture in some general way is
applicable only to variants of the conjecture which have been
ruled out by the counterexamples.  Thus, our ultimate conclusion
is that the evidence for the conjecture in any of its forms is
rather weak. If the bound is correct, it appears that this would
have to be due to some deep physics (for instance, due to some
aspects of quantum gravity as suggested in ref.~\cite{bekenstein},
or the string- or M- theory underlying the field theories used to
describe nature) beyond the frameworks of quantum mechanics and
quantum field theory.

We summarize our conclusions in Sec.~\ref{sect:summary}. We
relegate a number of the computational details to various
appendices.

\section{Evidence for the KSS bound}
\label{sect:evidence}

Before we begin, it is useful to briefly review the arguments of
KSS that have led to their proposed bound. The argument makes use
of the AdS/CFT duality from string theory
\cite{maldacena,gkp,witten,agmoo}. It is argued that in higher
dimensional gravity theories, black branes (higher-dimensional
analogs of black holes) have  finite temperature field theory
duals (specifically, $\mathcal{N} = 4$ supersymmetric Yang-Mills
theories at large $N_c$ and infinite 't Hooft coupling $g^2 N_c$)
that possess hydrodynamic properties such as viscosity. These
hydrodynamic properties can be related to gravitational properties
of the black branes, and the correspondence can be used to compute
transport properties \cite{KSS1}.  Using these methods the ratio
$\eta/s$ can be computed. A number of theories in this class have
been studied in the large $N_c$ limit at infinite `t Hooft
coupling. All of them have saturated the inequality of
Eq.~(\ref{eq:bound}) \cite{KSS1}. A general argument has been
given that all theories in this class at large $N_c$ and infinite
`t Hooft coupling must saturate the bound \cite{KSS2}.  Moreover,
one generally expects that as one weakens the coupling of an
interacting system, the viscosity should increase. One might,
therefore, expect that as the `t Hooft coupling is decreased from
infinity, the ratio $\eta/s$ should increase.  This has been seen
in an explicit calculation for the first correction due to finite
`t Hooft coupling for one particular theory \cite{FinitetHooft}.
Thus, it seems quite plausible that $\eta/s$ is bounded as in
Eq.~(\ref{eq:bound}), at least for those large $N_c$ field
theories which have super-gravity holographic duals.

The interesting question is whether the bound holds for some
general class of theories beyond this, and if so for which class
of theories.   Note that apart from the field-theoretic
calculations based on AdS/CFT, there is no reliable method to
calculate $\eta/s$ for {\it any} strongly coupled quantum fluid,
yet it is this class of fluids for which one expects the smallest
values of $\eta/s$. The optimistic view is that there could exist
a very general property of some large class of quantum fluids;
namely, the $\eta/s$ bound, which was unnoticed prior to the
AdS/CFT calculations in large measure because there was no
tractable way to compute the entropy and viscosity properties for
strongly coupled theories.   Of course, nature itself is an
excellent analog computer, and one way to probe whether there is a
bound which applies to the class of theories that describe the
real world is to ask whether there are any known fluids which
violate the putative bound.  In Ref.~\cite{KSS2}, KSS examined a
number of real life fluids, including liquid helium, liquid
nitrogen, and water, under a variety of conditions and found no
examples where the bound was violated. Typically, the ratio
$\eta/s$ for these fluids was found to be orders of magnitude
larger than the bound. This empirical data appears to be one of
the strongest pieces of evidence for the existence of a bound.

Additionally, a more heuristic argument can be made for the
existence of a bound \cite{KSS2,gsz}. Consider a relatively dilute
fluid which for simplicity is composed of one type of particle. By
dilute we mean: i) that the dynamics of the system is dominated by
two-body scattering, and ii) that the mean-free path $l$ between
collisions is much larger than both the thermal wavelength
$\lambda_{T}$ of the system and the characteristic range of the
interaction.  In effect, this dilute regime is weakly coupled from
the point of many-body physics; quantum many-body effects are
unimportant. This is the regime which can be accurately described
via a Boltzmann equation \cite{liboff}.  A simple kinetic theory
estimate of the shear viscosity in this regime was derived long
ago by Maxwell:
\begin{equation}
\eta \sim n p_T l \sim \frac{p_T}{\sigma}
\end{equation}
where $n$ is the density and $p_{T} = 2 \pi/\lambda_T$ is the
thermal momentum, and we have used the dilute-gas relation $n
\sigma l \sim 1$, where $\sigma$ is the scattering cross-section
at thermal energies \cite{liboff}.  (For a nonrelativistic system
$p_{T} \sim (m T)^{1/2}$, while for a relativistic system $p_{T}
\sim T$.) In the dilute regime, the entropy density is well
approximated by the free gas entropy density, and up to
logarithmic corrections in $m$ and $T$ the entropy density $s$ is
just proportion to the density. Combining these relations allows
us to write the ration $\eta/s$ as
\begin{equation}
\frac{\eta}{s} \sim \frac{p_T}{ n \sigma} \;
.\label{handwave}\end{equation}

Clearly, the expression for $\eta/s$ in   Eq.~(\ref{handwave})  is
monotonically decreasing with $n$. At first glance one might think
that by simply increasing $n$ one can reduce $\eta/s$ to as small
a value as one likes. However, Eq.~(\ref{handwave}) is only a
useful estimate in the dilute limit.  Increasing the density, the
mean free path shrinks, and eventually becomes comparable to
either the range of the interaction or the thermal wavelength.
Beyond this point, quantum effects alter the analysis, and one
enters a strongly coupled regime.  Presumably, these quantum
many-body effects cause the ratio of $\eta/s$ to stop decreasing
and begin increasing. From these simple scaling arguments, it is
easy to see that the density for $l \sim \lambda_T$ occurs when
$\eta /s \sim 1$. Therefore, at the length scale for which the
quantum many-body effects are expected to begin to increase the
ratio $\eta/s$, the effective minimum (and hence the lower bound)
of $\eta/s$ is on the order of 1.

Such general scaling and uncertainty arguments suggest that for
any given system the minimum value of $\eta/s$ will likely be of
order unity (or larger if the thermal wavelength is shorter than
the range of the interaction). This argument is heuristic and
does not explain why the number of order unity should be $(4
\pi)^{-1}$, but it is certainly consistent with it.  A somewhat
more sophisticated version of this argument may be found in
Ref.~\cite{gsz}.

\section{Classification of the variants of the KSS conjecture}
\label{sect:classification}

To discuss the various versions of the KSS conjecture
systematically, it is useful to classify the possible domains of
validity of the bound.  In doing so we focus on two
distinguishable aspects of the domains of validity. The first
aspect is the type of theory for which the conjecture is supposed
to apply. The bound was originally found in a very limited class
of theories --- large $N_c$ gauge theories with super-gravity
duals --- and assumed to hold for a broader class of theories.
Thus, the first matter that we need to characterize are the
classes of theories for which the bound may hold. The second
aspect to be characterized is the degree of stability of a fluid
described by some given class of underlying dynamical theory.  In
particular, this second classification delineates whether the
bound is to be taken to hold for stable fluids only, or for
long-lived metastable fluids as well.

Table I outlines a set of possible categories for both of the
above aspects of the domain of validity for the bound. The listing
of theory classes is intended to be ordered, more or less, in
decreasing scope: {\it i.e.,} as one descends the list, the
possible number of fluids which can be described by each
subsequent set of theories decreases.

\begin{table}[htb]
\renewcommand{\theenumi}{\Roman{enumi}}
\renewcommand{\theenumii}{\arabic{enumii}}
\renewcommand{\labelenumii}{\theenumii.}
\renewcommand{\labelenumiii}{3$^\prime$.}
\begin{enumerate} \item{Class of Underlying Theories}
\begin{enumerate}
    \item Any quantum mechanical system.
    \item Any nonrelativistic quantum mechanical system with one
    component of spin 0 or 1/2.
    \item Any ``sensible'' quantum field theory.
    \begin{enumerate}
    \item Any ``sensible'' quantum field theory with $\mu = 0$.
    \end{enumerate}
\end{enumerate}
\item{Stability Class of Fluids}
\renewcommand{\theenumii}{\alph{enumii}}
\begin{enumerate}
    \item{Absolutely stable fluids only}
    \item{Metastable and stable fluids}
\end{enumerate}
\end{enumerate}
\caption{Classification of the many forms of the conjectured bound
for the ratio of shear viscosity to entropy density.}
\end{table}

Using the classifications delineated in Table I, each variant of
the conjecture can be labelled by a pair of characters, one chosen
from the list of classes of underlying theories, and another
chosen from the list of stability classes. For example, if one
takes the original form of the conjecture that the bound applies
to ``all fluids'' to mean that it applies to all fluids described
by quantum mechanics, then the conjecture is of class 1a or 1b,
depending on whether one wishes to restrict the domain of validity
to absolutely stable fluids or not. Note that the list of theory
classes described in Table I may not be an exhaustive one, but it
is intended to include the natural interpretations of previously
published variants of the conjecture and some modest extensions
thereof.

In the next two subsections, we will further examine the classes
of theories and the fluid stability classes to which the bound
might apply. The first subsection will discuss the applicability
of the conjecture to the various classes of the theories that we
have delineated in Table I.  The second subsection will explore
the issue of stable versus metastable fluids.  After that, we will
discuss the applicability of the various versions of the
conjecture to different realistic fluids.

\subsection{Classes of Theories}
\label{sect:classificationTheory}

A fluid can be described theoretically as a many-body system whose
constituent particles are mobile enough to sample the complete
position space of the fluid.  We can define a ``theoretical fluid"
by defining the interactions between particles that make up the
fluid. Of course, real fluids may be regarded as theoretical
fluids as well --- they are the theoretical fluids associated with
the correct theory of nature.  The logic of the KSS conjecture is
that the $\eta/s$ bound, which was discovered in the context of
gauge theories with super-gravity duals, applies to a broader class
of theories.
Part I of Table I lists a number of possible classes of theories
for which the KSS bound might be taken to apply.

The list of classes of theories may seem somewhat peculiar.  It
was generated in part to reflect the possible ways to interpret
the variants of the KSS conjecture on the market.  There is
another reason to consider these classes. In many ways, the most
natural class of theory to consider is class $1$, the general
class of systems describable by quantum mechanics.  All of the
heuristic arguments in Sec.~\ref{sect:evidence} in support of a
generalization of the KSS conjecture to theories beyond those
described by AdS/CFT at large $N_c$ apply if the generalization is
to generic quantum mechanical systems.  (As we will see later,
this is not true of any of the alternatives.) However, it is easy
to see (by explicitly constructing counter-examples) that this
variant in its full generality {\it cannot} be correct.  The other
classes of theories in Table I may be thought of as ways to
restrict these classes of theories to which the bound should apply in order
to evade the problems with class 1.

As was briefly noted early on by KSS \cite{KSS2} and subsequently
addressed in more detail in Refs.~\cite{cohen} and \cite{spain},
the bound may be violated by considering a nonrelativistic fluid
composed of an extremely large number of distinct species, which
are all degenerate in mass, and interact with each other via
identical interactions. The key point is that by increasing the
number of species while keeping the total density of particles and
temperature fixed, the shear viscosity $\eta$ is left essentially
the same as in a single species fluid, while the entropy grows
through the Gibbs mixing entropy.  By making the number of species
exponentially large, the bound can be violated. A detailed
discussion of how this works is given in Sec.~\ref{sect:class1}.

The variants of the conjecture considered by KSS in
Ref.~\cite{KSS2} are essentially those in classes $2$ and $3'$.
These evade the problem of Gibbs mixing entropy in very
different ways. Class $2$ does this by explicitly limiting the
number of species in the fluid to no more than 2 (the number of spin states
of a spin-1/2 system), and thereby appears to restrict the growth in
Gibbs entropy.  Class $3'$ does this indirectly by restricting the
chemical potentials to zero: with a zero chemical potential and a
quantum field theoretic system, one cannot independently adjust
the density of each of the particle species, since each particle
density is fixed by the temperature and masses of the particle.
Thus by adding species at a fixed temperature, one necessarily
changes the total density of particles.

There is a subtlety associated with the theories of class $2$. The
issue concerns the precise definition of a fluid with ``one
component.'' Suppose that we have a many-body system with one type
of particle, $A$, which interacts through some two-body potential.
Suppose further that this interaction is attractive and some
number of two-body (and/or many-body) bound states (molecules)
exist.  One might wish to regard a fluid composed of particles of
type $A$ as a fluid with one component, since ultimately
everything in the fluid is composed of one type of particle.
However, the kinetic degrees of freedom whose motion describes the
fluid include both the atoms and the molecules, and the system is
effectively a multi-component fluid. Furthermore, one may naively
suggest that fluids which are composed from only one type of
molecule, such as water, may be considered as having a single
species. However, water molecules (and many other molecules) have
rotational and vibrational excitations which can be accessible.
These excitations cause the fluid to act like a multi-species
fluid with each excited state behaving as a distinct species. For
the purposes of the discussion here, we will therefore consider a
system to be of  ``one component'' only if either of the
following two conditions are met:  first, the particles making up
the fluid do not form bound states, and second, the internal
excitation energies of the particles are sufficiently high so
that the excited states are not populated due to the temperature.
Otherwise, we will consider the system to be a multi-component
fluid.

Variants associated with class $3$, which limit the conjectured
bound's applicability to systems described by ``sensible quantum
field theories," attempt to evade the entropy problem in a more
subtle way.  The modifier ``sensible" was introduced in this
context  by Son and Starinets \cite{SS1}.  In this context, the
term ``sensible" might be taken as a synonym for ``well defined"
--- that is, a quantum field theory in which, at least in
principle, all observables may be calculated without additional
{\it ad hoc} input.  From a practical perspective, the term
``sensible'' may be taken to refer to UV-complete theories;
namely, those which are sensible down to arbitrarily short
distances, and thus do not require additional prescriptions for
dealing with uncontrolled short-distance physics.  Thus,
``sensible quantum field theories'' would be taken to include
asymptotically free field theories such as QCD or conformal field
theories.  It should be noted here that the classes of theories
believed to be UV-complete are rather limited. Many renormalizable
theories with which we have considerable experience are probably
not ``sensible" (at least perturbatively) in the sense used
here.  For example, theories such as QED and linear sigma models
are presumably not ``sensible'' in that it is generally thought
that unless they are trivial, they are likely to be ill defined in
the ultraviolet.

How might the restriction to ``sensible'' quantum field theories
possibly evade the difficulty posed by Gibbs mixing entropy?
Recall that the violation of the bound may well require an
extremely large number of essentially identical species of
nonrelativistic particles.  Accordingly, it is difficult to find
any realistic situation where it occurs for real world fluids
\footnote{Note, however, that Ref.~\cite{spain} explores the
possibility that it may be possible to violate the KSS bound with
a gas of fullerenes}. One might hope that the difficulty of
constructing practical examples of such fluids might actually
reflect some deep and previously undiscovered principle. This
hypothetical principle must go beyond that which is contained
implicitly in quantum mechanics, since quantum mechanical systems
can be found which violate the bound in Eq.~(\ref{eq:bound}).
Thus, it is natural to ask if such a principle could have a
quantum field theoretic origin.  This gains some credence from the
fact that the conjectured bound was first seen in a particular
class of ``sensible" quantum field theories (conformal field
theories with gravity duals).  Thus, one might speculate that the
bound should only apply to systems which are ultimately described
by sensible quantum field theories, and therefore it should not be
possible to find a UV-complete field theory that can give rise to
a system that can violate the KSS bound.

On its face, it seems quite implausible that constraining the
relativistic field theory underlying a non-relativistic fluid to
be UV-complete should somehow rule out nonrelativistic fluids of
many components which violate the bound in Eq.~(\ref{eq:bound})
through a very large Gibbs mixing entropy. After all, the short
distance dynamics of the underlying quantum field theory typically
occur on radically different scales than the scales of the
effective degrees of freedom in the nonrelativistic gases of
interest. Accordingly, it is very difficult to see how a
constraint on the dynamics on $\eta/s$ for the fluid can arise
naturally.  Moreover, as noted in ref. \cite{KSS1}, even after
units are restored the speed of light does not appear in the
bound.  Thus, it is very hard to understand how the origin of the
bound could be related to the relativistic nature of the
underlying field theory.

The preceding arguments suggest that it is very hard understand
from first principles why a restriction to ``sensible''
relativistic field theories ought to yield the bound. However,
naive attempts to increase the number of nonrelativistic species
of particles in a gas by increasing the number of types of
particles in the underlying quantum field theory can easily cause
a theory to lose asymptotic freedom and thereby ceasing to be a
``sensible'' quantum field theory \cite{chen2,cohen}.

It is useful to illustrate how this can happen.   Let us consider
a nonrelativistic gas which is predominantly composed of one type
of pion of mass $m_{\pi}$,  for instance the  $\pi^+$. Such a gas
undoubtedly has its origins in QCD, a UV-complete quantum field
theory.  To describe such a gas in the context of QCD, we can
consider the theory at a finite temperature $T$, and a chemical
potential $\mu_u$ for the up quark $u$ of the form $\mu_u
\overline{u} \gamma_0 u$. (It is unnecessary to also impose a
chemical potential for the down quarks.) If the system is in the
regime $T \ll m_{\pi}$ and $\Lambda \gg \mu_u > m_{\pi}$ where
$\Lambda$ is a typical hadronic scale of order 1 GeV, then it is
essentially a nonrelativistic gas of $\pi^+$ mesons.  Now suppose
that we wish to generalize this to a many-species pion gas. To do
this, let us generalize QCD to include $N_f$ degenerate flavors of
quarks with $N_f$ large and even. Suppose we add a common chemical
potential $\mu_c$ for half of the flavors:
\begin{equation}
\sum_{j=1}^{N_{f}/2} \, \mu_c \overline{q}_j \gamma_0 q_j
 \end{equation}
while keeping $T \ll m_{\pi}$.  This will create a
nonrelativistic system containing $N_f^2/4$ types of pions (each
one with a quark of type $q_j$ with $j\le N_f/2$ and an anti-quark
of type $\overline{q}_k$ with $k > N_f/2$).  By carefully tuning
$\mu_c$ while increasing $N_f$, the total density of pions can be
kept fixed while increasing the number of species.  This appears
to allow one to create the conditions in which the Gibbs entropy
dominates the ratio of $\eta/s$ and causes a violation of the KSS
bound.

However, there is a catch.  Recall that for small $g$, the beta
function for QCD is given by
\begin{equation}
\beta(g) = - \frac{g^3}{16 \pi^2} \left( \frac{11 N_c}{3} - \frac{2
N_f}{3} \right ) \; .
 \end{equation}
Asymptotic freedom requires that $11 N_c > 2 N_f$.  By increasing
$N_f$ in order to violate the bound in Eq.~(\ref{eq:bound}), the
underlying theory is pushed outside of the domain of ``sensible''
theories. Of course, one might try to evade this by increasing
$N_c$ at the same time as one increases $N_f$; by fixing the ratio
$N_c/N_f$ as the large $N_f$ limit is taken, asymptotic freedom
can be maintained. However, recall that the cross section for
$\pi-\pi$ scattering scales as $1/N_c^2 \sim 1/N_f^2$
\cite{Wit}.  For a weakly interacting fluid, the shear viscosity
is expected to scale with the inverse of the cross-section
\cite{liboff}. Thus, by increasing $N_c$ along with $N_f$ to
maintain asymptotic freedom and keep the theory sensible, one
finds that $\eta \sim N_f^{2}$. On the other hand, the Gibbs
mixing entropy grows only with $\log (N_f)$, so $\eta/s \sim
N_f^2/\log (N_f)$ for large $N_f$.  As a result, in a pion gas in
the large number of species limit, the decrease in the cross
section associated with the $N_c$ scaling necessary to maintain
asymptotic freedom overwhelms the increase in Gibbs mixing
entropy due the to the $N_f$ scaling, and $\eta/s$ is driven to
infinity in the combined $N_f \sim N_c \rightarrow \infty$ limit.

The example of pion gases in QCD shows how the restriction to a
``sensible'' theory can prevent the system from ever getting into
a regime where the Gibbs mixing entropy dominates the ratio of
$\eta/s$ and thus violates the KSS bound.  The central question
underlying the theories associated with class $3$ is whether the
situation seen for pion gases in QCD is paradigmatic for all
sensible theories.

There is an additional important subtlety associated with the
notion of ``sensible'' in class $3$; namely, whether the standard
model should be regarded as a sensible quantum field theory.  The
standard model contains scalar fields and is probably {\it not}
UV-complete. This implies that class 3 should not apply to the
standard model {\it per se}.  However, the standard model may be
regarded as the low energy effective theory for some theory (a
field theory, a string theory, or something else) which {\it must}
make sense in the ultraviolet since it describes nature.  Thus, it
might be useful to regard ``the standard model,'' as described in
the textbooks, to include the appropriate UV-completion for real
world situations, and hence be ``sensible.''

Having delineated some of the possible domains of validity of the
conjectured bound on $\eta/s$, in the next subsection we will
discuss the matter to which stability classes of fluids the
conjectured bound may apply.

\subsection{Metastability}
\label{sect:classificationMeta}

In addition to distinguishing variants of the conjecture according
to the classes of underlying theories to which they apply, we also
need to discuss the stability classes of the fluids for which the
$\eta/s$ bound may apply. A fluid can be described as either
stable or metastable. In this subsection, we will examine some of
the issues associated with the applicability of the bound to
stable and metastable fluids.

We are defining a metastable fluid to be one which is in a
macroscopic state which is not the state of lowest free energy;
such a fluid is expected to decay over time to the true
macroscopic ground state.  If the time scale of the decay is
extremely long compared to other relevant time scales, the fluid
is considered to be metastable.  A stable fluid, on the other
hand, is one in which no decay is possible, {\it i.e.}, the fluid
is in its ground state and will remain there unless it is
perturbed. Metastable fluids are characterized by at least two
relevant time scales. First, there is $\tau_{\text{fl}}$, which is
the longest microscopic time scale relevant for fluid motion.  In
practice, for a typical real world fluid, $\tau_{\text{fl}}$ might
be taken to be several times the characteristic collision time
between molecules.  Thus, $\tau_{\text{fl}}$ characterizes the
minimum time scale for which it is meaningful to talk about
macroscopic fluid behavior.  Next, there is the time scale
$\tau_{\text{meta}}$ for the decay from a metastable fluid to a
stable (that is, lowest-energy) configuration.

The characterization of the fluid clearly depends on the ratio
$\tau_{\text{fl}}/\tau_{\text{meta}}$.  If
$\tau_{\text{meta}}/\tau_{\text{fl}} \sim 1$ or less, then the
decay time is of the same order or less than the characteristic
time for fluid-like behavior, and it is not meaningful to describe
the system as being in a well-characterized fluid state.  In
effect, in such a regime the fluid is so unstable that one cannot
measure properties like shear viscosity or entropy before the
system decays into a qualitatively different type of fluid.
However, if $\tau_{\text{meta}}/\tau_{\text{fl}} \gg 1$, the decay
time scale is much longer than the time scale of the measurements
needed to determine fluid properties such as the shear viscosity.
In this case, the fluid can be said to be metastable, and
properties such as viscosity and entropy are essentially well
defined in the metastable phase. For an extremely large
$\tau_{\text{meta}}$, the metastable fluid acts to a very good
approximation as if it were a stable fluid.  We should note that
many systems which we obviously characterize as fluids in the real
world are actually metastable. An extreme example is nitroglycerin
($C_3H_5(NO_3)_3$).  Above its melting point of $13.2^\circ$C it
is clearly a fluid --- it will slosh around in a beaker.  However,
liquid nitroglycerin is obviously {\it not} in a configuration at
the minimum of the free energy --- considerable energy can be
released when the molecules break up and rearrange. It is
noteworthy that in the real world, $\tau_{\text{meta}}$  for
metastable fluids is typically {\it many} orders of magnitude
larger than  $\tau_{\text{fl}}$.

There are two ways in which a system can be metastable in the
sense used here.  The first is the rather typical example in
statistical physics in which a macroscopic phase is locally stable
while being globally unstable.  That is, any small fluctuation of
a macroscopic fluid property ({\it e.g.}, density) from its value
in the metastable phase increases the free energy, but large
fluctuations can lower it.  This is quite familiar in systems
which can undergo first-order phase transitions.  The system can
be beyond the phase transition point but stay in the old phase.
Thus, for example, water may be supercooled or the relative
humidity can be greater than 100\%.   Such systems can live for a
very long time (if undisturbed) since there is barrier which must
be either surmounted via thermal fluctuation or tunnelled through
quantum mechanically. In either case, if the barriers are large,
the lifetimes of the metastable phases grow exponentially.

There is a second way for a system to be metastable. A system can
be locally unstable in terms of {\it some} thermodynamic
variables, but the time scale associated with the local
instability can be very long.  It is this sort of metastability
which is relevant for many of the discussions in this paper. For
example, this can happen in chemical systems.  A system can be in
thermal equilibrium kinetically but not chemically;  however, the
time scale for reaching chemical equilibrium can be very large.
Suppose, for example, that one initially has a gas composed of
molecules of one type, $A$. Suppose further that the reaction $A +
A \rightarrow B + C$ (where $B$ and $C$ are two other types of
molecules) is exothermic, but the reaction rate is {\it very}
small compared to the rate of elastic scattering of particles of
type $A$.  This will happen if the activation energy for the
reaction is well above the temperature. In such a case, over very
long time scales the system will act like a fluid of molecules of
type $A$ in thermal equilibrium kinetically, despite being out of
thermal equilibrium chemically. Locally, as well as globally, the
system is not at a minimum of the free energy for all of the
thermodynamic degrees of freedom, but nonetheless behaves like a
fluid.

We noted above that when $\tau_{\text{meta}} \gg
\tau_{\text{fl}}$, fluid properties such as shear viscosity  are
essentially well defined. In a strict sense, however, they are
not.  As a matter of principle, transport properties, such as
shear viscosity, describe the linear response of a fluid to a
perturbation.  This response is dynamical, and takes a certain
characteristic time to play out. We can identify this time as
$\tau_{\text{fl}}$.  The transport properties are only well
defined to the extent that the underlying fluid does not change
its nature over this dynamical time scale. Since a metastable
fluid {\it does} change its properties over time, there is an
intrinsic ambiguity in any evaluation of $\eta$. One might expect
that any uncertainty in the value of $\eta$ is roughly of
relative order $\tau_{\text{fl}}/\tau_{\text{meta}} $.
Fortunately, in a good metastable system this is an exceptionally
small number, and the ambiguity is very small.

The issue of metastable fluids is important in the context of the
KSS conjecture.  The central question is whether the conjectured
bound applies to metastable fluids as well as to stable fluids.
This may seem like a relatively minor issue if the bound applies
to the theories in class $1$.  Then the question of whether the
bound applies to metastable fluids reduces to the issue of whether
it applies to normal stable fluids such as water, or whether it
also applies to metastable fluids such as nitroglycerine. However,
as will become apparent in the next subsection, if the bound only
applies to theories in class $3$, the question of whether the
bound applies to metastable fluids determines the bound
applicability to familiar real-world fluids.

To the extent that the KSS conjecture somehow captures an essential
property that a system needs to possess to behave as a fluid, one might
naturally assume that it should also apply to metastable systems
whose macroscopic behavior is clearly that of a fluid.  There is
an objection of principle that could be made here, in that the
conjecture is sharp --- it provides a definite bound for
$\eta/s$ --- while the quantities $\eta$ and $s$ are intrinsically ambiguous for a metastable fluid.  Of
course, as noted above the ambiguities are very small for
long-lived metastable fluids.  Accordingly, it is highly plausible
that the conjecture, if correct, applies to metastable fluids with
one minor alteration: the bound may be slightly violated, but all
possible violations must be within the scales of ambiguities of
the quantities.  In practice, for real metastable fluids, these violations are
extraordinarily small, and as a practical matter the bound would
then be taken to hold for any long-lived metastable fluid.   We
generally take the view that is unnatural for there to be a
fundamental property which applies to all stable fluids, but which
does not apply --- even approximately --- to metastable fluids no matter
how long-lived.  It seems far more natural to assume that in the
limit of infinite lifetime, a metastable fluid would be
indistinguishable from a stable one, and that it would share all of the essential
properties of stable fluids.  Having said this, as a logical
matter it is certainly possible that the bound only applies --- even
approximately --- only to absolutely stable fluids.  Accordingly, it
is important to classify the possible $\eta/s$ conjectures according to whether or
not they apply to metastable systems.

\subsection{Applicability of the various classes to real fluids}
\label{sect:classificationApplicability}

Having enumerated various forms of the conjecture, it is important
to see the types of realistic fluids to which they apply.  In
Table II, we show the applicability of the various forms of the
conjecture to four different types of fluids which serve to
illustrate the broad issues of where the various classes apply.
The fluids we examine --- the quark-gluon plasma, liquid helium,
water, and nitroglycerine
--- were chosen to serve as paradigms for broad classes of fluids.

\begin{table}[htb]
\begin{tabular}{c|c|c|c|c}
Variant & QGP & $He$ &$H_2O$ & $C_3H_5(NO_3)_3$ \\
\hline
1a. & Y & Y & Y & N \\
1b. & Y & Y&  Y & Y \\
2a. & N & Y & N &  N \\
2b. & N & Y & N &  N\\
3a. & Y & N & N & N \\
3b. & Y & Y & Y & Y \\
$3^{\prime}$. & Y & N & N & N \\
 \hline
\end{tabular}
\caption{Table showing if each variant of the conjecture can be
applied (at least approximately) to either the quark gluon plasma
(QGP), liquid helium ($He$), water ($H_2O$),liquid nitroglycerin
($C_3H_5(NO_3)_3$); Y(es), N(o)} \label{tab:conjecture}
\end{table}

First, consider the quark-gluon plasma.  It is generally believed
that the dynamics of high energy heavy ion collisions depend
essentially on QCD alone --- {\it i.e.,} electroweak effects are
small.  Moreover, it is generally thought that the system
thermalizes, at least approximately, over reasonably large spatial
regions, and that in these regions the net baryon density is low
since the bulk of the baryon number goes down the beam pipe. Thus,
to a good  approximation these regions are well described by QCD
at finite temperature and zero chemical potential.  If the
temperature in these regions is large enough (above $\sim$ 170
MeV), these regions can be said to contain a quark-gluon plasma.
Note that in saying this we do not necessarily imply that QCD has
undergone a phase transition into a quark-gluon plasma phase; a
rapid cross over into a qualitatively high-temperature regime is
adequate.

Clearly, nontrivial approximations are needed in order to
describe this physical system in terms of thermalized QCD at zero
chemical potential.  However, if one accepts these approximations
as valid --- as we will do implicitly for the purpose of this
discussion --- then one has a well-characterized field theoretic
description of the quark-gluon plasma.  Within that
characterization, it is clear that all of the variants of the
conjecture  should apply to this system, except for the variant
with theories of class $2$.  As a quantum field theory, it is
certainly a quantum mechanical system, and thus falls neatly into
class $1$.  QCD is the archetypical example of a ``sensible''
field theory: it is asymptotically free, and hence is described by
theories of class $3$.  Moreover, as a system at zero chemical
potential, it falls into class $3'$.  Clearly, since the
quark-gluon plasma is a relativistic system with many components,
it does not fit into class $2$.

Next, consider liquid water, which is truly an archetypical
example of a fluid. Clearly from the perspective of chemical
interactions, water is a stable fluid.  One could model water to
very high accuracy using a many-body quantum-mechanical
description based on electrons, oxygen nuclei and hydrogen nuclei
as the basic degrees of freedom, interacting via a Coulomb
potential and (small) magnetic moment interactions. While in
practice, it would be very hard to compute $\eta/s$ from such a
model, in principle it {\it is} computable, and we have every
reason to believe that such a description would be very accurate.
Thus, one expects that variants $1$a and $1$b of the conjecture
should apply to water.

However, one does not expect variants of class $2$ to apply.  The
previous description based on electrons and nuclei clearly
violates the condition that there is only one component to the
fluid.  One might try to avoid this by considering an effective
quantum mechanical model of the dynamics of water where the
fundamental building blocks are water molecules interacting via
effective interactions.  Such a description would be under the
umbrella of class $2$ provided that only a single internal quantum
state of the water molecule was relevant to the dynamics. However,
the minimum excitation energy of a water molecule is 0.16 K
\cite{waterrot}, which corresponds to a rotational level, while
for liquid water $T > 273$K. Thus, in practice, water molecules
are not to be found predominantly in their lowest energy level in
liquid water
--- many rotation levels of the molecules are excited, and the
system does not act like a single-component fluid.

The applicability of conjectures based on theories of class $3$
(``sensible'' quantum field theories) to water is subtle and
perhaps somewhat counterintuitive.  Since water is a real world
fluid and thus is presumably described by the standard model --- a
quantum field theory --- it seems natural that water be included
within the variants of the conjecture based on class $3$.  As
noted above, there is some question as to whether we should
consider the standard model to be a ``sensible'' quantum field
theory, but for the moment let us assume that it is legitimate to
do so.  With this assumption, it may seem obvious that variant
$3$a applies to water since water is a stable fluid.  However,
this is is not the case.  Although it is stable chemically, water
is {\it not} stable under the dynamics of the standard model:
nuclear reactions are part of the standard model and can alter the
constituents of water. For example, it is energetically allowable
for two of the hydrogen nuclei in water to fuse in the reaction $p
+ p \rightarrow d + e^+ + \nu_e$. Of course, the decay time of
nuclear fusion in water is {\it very} long, indeed much longer
than a Hubble time.  The reason for this is simply that the
Coulomb barrier is very large compared to thermal energies, and
the rate of thermal fusion is thus exponentially small. Thus from
the perspective of the standard model water is metastable rather
than stable: variant $3$a of the conjecture does not apply to
water, but variant $3$b does, at least to the extent that we can
consider the standard model, and whatever lies beyond it, as a
sensible quantum field theory. Clearly, theories of class $3'$ do
not apply to water since this class is a subclass of $3$a.

The fact that water is not stable under the dynamics of the
standard model reflects the conservation laws of the standard
model.  Clearly, under the standard model the number of hydrogen
and oxygen nuclei do not represent conserved quantities. Apart
from electric charge, the only global conserved quantity in the
standard model is $B-L$; due to anomalies the baryon number $B$
and lepton number $L$ are not separately conserved.  Thus, the
only type of stable fluid we can specify in the standard model is
one with a fixed chemical potential for $B-L$.

One might argue that the rates of nuclear reactions are so slow that
they could not possibly be relevant to the validity of the
conjecture. While this is a very plausible argument, it is simply
an argument against a requirement that the conjectured bound needs absolute
stability rather than metastability.

There is an alternative argument which can be made that variant
$3$a can apply to water \cite{Son}.  Nuclear reactions are totally
irrelevant at the scale of interest for water.  Thus, to study
water one might replace the standard model with a variant of
quantum electrodynamics containing electrons and fundamental
fields representing the proton and the oxygen nucleus.  To the
extent that hyperfine effects involving the nuclear spins are
unimportant to the dynamics of water, such a system will behave
like water and will be absolutely stable, apparently putting water
in the domain of variant $3$a. However, there is a problem with
this setup: QED is not asymptotically free and as a result it is
presumably not ``sensible''.   One might hope to evade this by
embedding this low energy QED-like theory into another theory
which a) {\it is} asymptotically free, b) leaves the low-energy
QED physics essentially unaltered, and c) does not introduce any
instabilities for water.  Unfortunately, it is by no means clear
that it is possible to find any field theories which meet these
criteria. Until such a theory is constructed, we will take the
view that variant $3$a should not be regarded as applying to
water.

Other real world fluids dominated by chemical ({\it i.e.,}
electromagnetic) interactions are similar to water in terms of
their classification, with obvious modifications. Thus, for
example, liquid helium is like water in being described by
variants 1a, 1b, 3b, and not 3a.  It differs in that the lowest
excitations for helium are electronic in nature, since helium
(unlike water) is an atomic fluid as opposed to a molecular fluid.
Since liquid helium temperatures are well below the excitation
energy for electronic transitions, the atoms in liquid helium are
essentially all found in their ground state.  Thus, it is possible
to model liquid helium with good accuracy in terms of a quantum
mechanical many-body system with fundamental helium atom degrees
of freedom interacting via an effective potential.  Within the
framework of such a model, liquid helium, unlike water, falls
within the domain conjectures of classes $2$a and $2$b. Similarly,
nitroglycerine is like water in terms of the variants of the
conjecture which describe it, with the exception of class $1$a
which describes water (which is a stable fluid chemically) but not
nitroglycerine (which is obviously metastable).

Having discussed a framework for labeling the possible variants of
the $\eta/s$ bound conjecture, in the next three sections we will
construct and discuss counterexamples to variants of the
conjecture of classes  1, 2, and 3.  We will ultimately show that
that only class 3a (and its subclass $3'$) remains viable.

\section{Class 1}
\label{sect:class1}

The first variant of the KSS conjecture that we will closely
examine is class $1$: the conjecture that $\eta/s \geq 1/4\pi$
for all fluids described by quantum mechanics.   This variant
seems to be very close to the original form of the bound proposed
by KSS \cite{KSS1}. Note that this variant of the conjecture has
much stronger support than the other variants; all of the
heuristic arguments as well as all of the empirical evidence
given in support of the KSS bound support this variant. This
variant has the widest applicability, as it it applies to any
fluid, both relativistic and nonrelativistic ones, and both
physically realizable or purely theoretical fluids provided they
are described by quantum mechanics.

However, Ref.~\cite{KSS2} and others \cite{cohen,spain} have noted
that this variant of the conjectured bound can be violated by
considering a fluid with a large number of different species. In
this section, we elaborate on the previous arguments of
Ref.~\cite{cohen} to describe a nonrelativistic quantum
mechanical system which violates the conjectured bound.

\subsection{A nonrelativistic gas}

Reference \cite{cohen} considers a nonrelativistic quantum
many-body system with a large number of species for which the
computation of the ratio $\eta/s$ is analytically tractable, up
to corrections which can be made arbitrarily small. By imposing a
particular set of scaling relations on the parameters of the
system, it is possible to demonstrate that $\eta/s$ can violate
variants 1a and 1b in the limit of a large number of species. We
review this argument here.

Consider a gas composed of a number ($N_s$) of distinct species of
spin-0 bosons of degenerate mass, $m$, which can interact via a
two-body potential. The two-body potential is identical for all
species, but is limited to a finite range, $R$. The gas is in
thermal equilibrium at a temperature $T$, and has the same density
for each species, $n_a = n/N_s$, where $n$ is the overall density
of the system. The system is in a low temperature and low density
regime such that
\begin{equation} \label{eq:regime}
R^{-2}, a^{-2} \gg mT \gg n^{2/3},
\end{equation}
where $a$ is the scattering length, and $m T$ is the thermal
momentum squared. This regime can be maintained by using the
following scaling of the density and temperature:
\begin{equation} \label{eq:scale1}
n = \frac{n_0}{\xi^4} \qquad T = \frac{T_0}{\xi^2},
\end{equation}
where $n_0$ and $T_0$ are independent of the dimensionless scaling parameter $\xi$.
With a sufficiently large value for $\xi$, Eq.~(\ref{eq:regime}) can be easily
satisfied.

In this density and temperature regime, the entropy for the system
is simply that of a classical ideal gas, with small corrections.
The key point is that the temperature is high enough relative to
$n_0^{2/3}/m$ for the classical expression to hold, while the
density is low enough to neglect the interactions. The entropy
density can then be written in terms of the scaling in
Eq.~(\ref{eq:scale1}) as
\begin{equation}
s \simeq n_0 \biggl(\log \Bigl( \frac{(m T_0)^{3/2}}{n_0}\Bigr)
+\frac{5}{2}+\log(\xi) + \log(N_s)\biggr),
\end{equation}
where the term $\log(N_s)$ is associated with the Gibbs mixing
entropy of the $N_s$ different species.

Furthermore, in this density and temperature regime, the thermal
wavelength is much shorter than the inter-particle spacing,
meaning that the many-body dynamics are essentially classical.
Moreover, the low density implies that the many-body dynamics are
dominated by binary collisions, implying that the system is in
the regime of validity for the Boltzmann equation \cite{liboff}.
The low temperature further implies that the two-body collisions
are dominated by s-wave scattering, with a cross section
essentially unchanged from its zero momentum value. That is,
two-body scattering in this system can be approximated as
isotropic and energy independent, which is formally the same as
classical hard sphere scattering.

The shear viscosity is analytically calculable in such a system
\cite{liboff}, and it is given by $\eta = C_{\text{hs}} \sqrt{m
T}/d^2$, where $d$ is the diameter of the hard spheres, and
$C_{\text{hs}} \approx .179$ is a coefficient that is numerically
calculable \cite{chs}. Identifying the scattering length $a$ as
the effective hard sphere diameter, we can now calculate the ratio
$\eta/s$:
\begin{equation} \label{eq:ratio1}
\frac{\eta}{s} \simeq \frac{ C_{\text{hs}} \xi^3 \sqrt{m T_0}}{a^2
n_0 \Bigl(\log \Bigl( \frac{(m T_0)^{3/2}}{n_0}\Bigr)
+\frac{5}{2}+\log(\xi) + \log(N_s)\Bigr)}.
\end{equation}
Corrections to Eq.~(\ref{eq:ratio1}) are suppressed by powers of
$1/\xi$ and should become irrelevant for sufficiently large $\xi$.

The derivation of Eq.~(\ref{eq:ratio1}) required the system to be
in a low density and low temperature regime such that a classical
approximation for both the shear viscosity and the entropy density
can be made. This limit does not place any constraints on the
number of species of particles in the fluid. Accordingly, one can
demand that the number of species scale exponentially with the
scaling parameter:
\begin{equation} \label{eq:scale2}
N_s = \exp (\xi^4)
\end{equation}
As the temperature and density decrease, the number of species increases
simultaneously. When Eqs.~(\ref{eq:ratio1}) and (\ref{eq:scale2})
are combined, the large $\xi$ scaling of the ratio is
\begin{equation} \label{eq:ratio2}
\frac{\eta}{s} \simeq \frac{1}{\xi} \frac{C_{\text{hs}} \sqrt{m
T_0}}{a ^2 n_0}
\end{equation}
up to power law corrections in $1/\xi$. Clearly, in this combined
limit, the ratio $\eta/s$ can violate the conjectured bound simply
by making $\xi$ sufficiently large. This violation stems
completely from the large Gibbs mixing entropy associated with the
exponentially large number of species.

\subsection{Stability}
\label{sect:class1:stability}

In this subsection, we will discuss the stability class of the
fluid that we have described above. The argument in the preceding
section does not depend on the interparticle potential and thus
will continue to hold for any choice of the interparticle
potential. If we choose the interparticle potential to be purely
repulsive, the particles making up the fluid cannot lower their
energies by forming bound states. Therefore, with this choice, the
system that we have described above is a stable fluid with an
arbitrarily small value of the the ratio $\eta/s$.  This is
sufficient to demonstrate that this system is a counterexample to
\emph{both} class 1a and 1b variants of the conjecture.

While with the system above we were free to choose the interaction
potential to be whatever we wanted, in some other situations this
is not possible.  In particular, in our discussion of systems of
class 3, in Sec.~\ref{sect:class3} we will find that the
interaction potential there will necessarily be an attractive one.
To see the implications on the stability of a fluid of an
interaction potential with some attractive regions in a simple
context, we will now discuss the consequences of choosing an
interparticle potential with some attractive regions for the
system in the previous section.

One might worry that with such a potential, the fluid could lower
its energy by forming bound states, or by ``clumping" together;
that is, by forming macroscropic regions of higher density where
the attraction is enhanced and the free energy is lowered. If
either situation is possible, the fluid would then be either
unstable or metastable. As discussed in
Sec.~\ref{sect:classificationMeta}, in order to distinguish
between these two cases, we need to compare $\tau_{\text{met}}$,
the characteristic time for the phase to change macroscopically,
with $\tau_{\text{fl}}$.  We can show that in our scaling regime
$\tau_{\text{meta}}/\tau_{\text{fl}}$ diverges as $\xi^{5}$ or
faster, ensuring that when $\xi$ is large the system is
metastable.

The  type of metastability with the decay mechanism which yields
the fastest possible decay parametrically is for systems which can
form two-body bound states. As is well known, in a
nonrelativistic gas three-body collisions are necessary to allow
the formation of two-body bound states due to energy and momentum
conservation. Therefore, the decay time $\tau_{\text{met}}$ scales
with the time between three-body collisions in the system. The
characteristic time scale of the fluid $\tau_{\text{fl}}$ scales
with the time scale for two-body collisions. Therefore the ratio
$\tau_{\text{meta}}/\tau_{\text{fl}}$ has roughly the same scaling
as $\tau_3/\tau_2$, where $\tau_3$ and $\tau_2$ are the three-body
and two-body collision time scales, respectively.

The time between two-body collisions is essentially just the mean free
time of particles in the fluid. The mean free time $\tau_{\text{mf}}$ is related to the mean free
path $l_{\text{mf}}$ by
\begin{equation}
\tau_{\text{mf}} = l_{\text{mf}}/v,
\end{equation}
where $v$ is the rms velocity of particles in the fluid. In dilute
classical gases the mean free path $l$ can be related to the
density and the interaction cross section,
\begin{equation}
n l_{\text{mf}} \, \sigma \sim 1.
\end{equation}
The rms velocity $v$ can be related to the thermal momentum
associated with the fluid: $m v \sim  \sqrt{m T}$, where $m$ is
the mass of the particle, and $T$ is the temperature of the
fluid. Combining these equations and the scaling relations of
Eq.~(\ref{eq:scale1}), we see that the mean free time scales like
\begin{equation}
\tau_{\text{mf}} = \frac{1}{n \sigma} \sqrt{\frac{m}{T}} \sim
\xi^5 \frac{1}{n_0 R^2} \sqrt{\frac{m}{T_0}},
\end{equation}
where we have used the relation $\sigma \sim R^2$, with $R$ being
the characteristic range of the interaction.

In addition to $\tau_{\text{mf}}$, we must examine the
characteristic time that two particles spend interacting during a
collision, $\tau_{\text{int}}$. Equation (\ref{eq:scale1}) implies
that scattering is at low momentum.  As a result,
$\tau_{\text{int}}$ does not scale with $\xi$, since it is
essentially a function of the details of the two-body potential
and does not depend on $v$. The fraction of the time between
two-body collisions during which the particles are interacting is
$f \sim \tau_{\text{int}}/\tau_2 \sim \xi^{-5}$.

To form a two-body bound state, a three-body collision is
necessary.  That is, while two particles are in the process of
interacting, a third particle must collide with them.  In terms of
the quantities defined previously, the time scale for such events
is simply $\tau_3 = \tau_2/f$.  As a result, we see that
$\tau_{\text{meta}}/\tau_{\text{fl}} \sim \tau_3/\tau_2 \sim
\xi^{5}$, as claimed above.  Other mechanisms take longer
parametrically: if the most rapid decay involves the formation of
an $N$-body state, an analogous calcuation yields
$\tau_{\text{meta}}/\tau_{\text{fl}} \sim \xi^{5 (N-1)}$.


To summarize, the  arguments in this section show that the variants
of the KSS bound of class $1$ can be violated by a fluid with a
large number of species. Depending on the choice of an interaction
potential, the fluid that we have described can be either stable
or metastable.  While the example used to demonstrate the violation of the bound is
highly artificial and unlikely to be realizable even approximately
in a real world setting, as a mathematical matter it is a
legitimate counterexample.  The implication is that the most
well-supported and most widely applicable variants of the
conjecture --- those of class $1$ --- are not tenable.


\section{Class 2}
\label{sect:class2}

In this section, we discuss the variants of the KSS conjecture of
class $2$. This form of the conjecture states that $\eta/s \geq
1/4\pi$ for all nonrelativistic fluids composed of a single
species of particle of spin-$0$ or spin-$1/2$. This variant of the
conjecture is essentially the one that was proposed by KSS in
Ref.~\cite{KSS2}.  By restricting the number of allowable species,
this variant of the conjecture attempts to avoid the problem with
the Gibbs mixing entropy that allowed the construction of a
counterexample to the variants of class $1$.

Note at the outset that the evidence in support of this class of
conjecture is quite limited.  The AdS/CFT duality arguments do not
apply.  Since these calculations were done in the large $N_c$
limit, it is hard to understand how they could justify a bound
that fails for a large number of species and only works when the
number of species is small enough.  Moreover, much of the
empirical evidence in favor of a KSS bound does not apply to
variants of this sort. The term ``single-species" in this context
refers to systems whose constituents are either elementary or are
in their ground state and do not access higher excited states.  As
a result, liquid water is not covered in this variant of the
conjecture: water molecules in a liquid state can access
rotational modes, making water  a multi-species fluid from this
perspective.  This limits the applicability of this variant of the
bound mostly to mono-atomic fluids, such as liquid helium. Since
the vast majority of real world fluids are not in this class, the
fact that no known violation of the bound exists for real fluids
provides only modest support for the bound.

In this section, we will investigate a counterexample to variants
of class 2. We will give an example of a stable
quantum-mechanical system composed of only one kind of spin-$0$
particle that can violate the KSS bound.  Since the counterexample
is for a stable fluid it appears to rule out both variants 2a and
2b.

To demonstrate that the existence of a class 2 system violates
the bound, we first define the system by choosing a particular
two-body interaction potential. The properties of the fluid in a
non-relativistic regime are determined by the interaction
potential along with the temperature and the density. The basic
idea is to construct a two-body interaction of finite range which
has an an extremely large number of two-body resonant states
right above threshold.  We show that the entropy for such a
system has a lower bound, which by a judicious choice of
parameters can be made arbitrarily large, even though there is
only a single species of particles making up the fluid. Finally we
argue that the shear viscosity of such a system is not expected to
become uncontrollably large as the parameters are adjusted to make
the entropy grow arbitrarily. Thus it appears that the ratio of
$\eta/s$ can be made arbitrarily small within this class of
theory.

\subsection{Constructing the System}
In this subsection, we define a single-species fluid composed of
identical, stable, spin-0 particles. These identical spin-0
particles are considered to be the fundamental particles of the
fluid.  We will choose a finite-range two-body interaction that
supports no bound states (two-body or many-body) while supporting
an arbitrary number of arbitrarily low-lying resonant states in
the scattering amplitude. The resonant states may be long-lived
(depending on the choice of parameters of the potential), but it
is important that they are indeed resonant states, and \emph{not}
bound states, so that there is no question that the fluid is of a
single species.

Before discussing a detailed form of interaction which can
generate this situation, it is important to note at the outset the
interaction will require an exceptional degree of fine-tuning. The
principal reason for this is that we require that the range of the
interaction remains fixed as we add resonances.  We impose this
requirement because we wish to keep the density of the fluid fixed
as we add resonances in order to avoid having many particles
simultaneously within the range of interaction.  This creates a
strong constraint in which we require an exponentially large
number of nearly degenerate s-wave resonances near threshold for a
system of fixed spatial extent. A useful way to envision making a
system of finite size with multiple nearly degenerate two-body
resonances is to start by constructing a system with numerous
nearly degenerate two-body s-wave bound states and then add a
repulsive potential to push them into the continuum.

However, it is not trivial to create a large number of nearly
degenerate bound states with the same quantum numbers due to level
repulsion.  One way to proceed is by using a central potential
which has numerous nested spherical-shell-shaped wells; we denote
the number of wells as $N$. Clearly, if the spatial size of the
interaction is kept fixed as one goes to a regime of large $N$
(as is needed to achieve many bound states), the width of each
well in the radial coordinate, $r$, must be very small. To
understand the tuning of parameters that is required, it is
easiest to start by considering a system with a single well at a
fixed position---with the position corresponding to the positions
of one of the nested wells. The parameters are picked such that
the single-well system has a single two-particle bound state.
This can be achieved by tuning either the width or the depth of
the well, or both.
%
Arbitrarily narrow wells can always be constructed to have a
single bound state with fixed binding energy by making the well
deep enough. In taking the width in the radial direction to be
small (as we are forced to), in essence one is fine-tuning the
depth of the potential, $V_0$, so that the binding energy is a
very small fraction of $V_0$.  For a generic well, it is not
possible to do this for more than one bound state level. The bound
state wave functions will be localized in the radial coordinate
around the well.  Note that there is a considerable level of
parameter-tuning necessary to achieve this.

Now suppose we consider a system with all $N$ of the wells present
simultaneously. The parameters would need to be further tuned so
that the bound states in each of the $N$ wells are nearly
degenerate. To the extent that bound state wave functions for the
single well case were well localized---{\it i.e.}, have a
spreading in $r$ which is much less than spacing between levels
--- the full system will have $N$ nearly degenerate bound states,
each with an energy near that of the single well case.  However,
if that condition is not met, there will be significant level
repulsion and the condition of near degeneracy will be destroyed.
The characteristic spread of the wave functions is $(m B)^{-1/2}$
where $m$ is the particle mass, and $B$ is the binding energy.
Accordingly, to include a large number of wells within a fixed
radius while keeping the levels nearly degenerate requires that
the binding energy be tuned to be large.

There is a final level of tuning required.  We have shown that
considerable tuning is required to get $N$ nearly degenerate
deeply bound states in a system with $N$ nested wells with fixed
range.  However, we wish to have a system with $N$ resonances.  We
can do this by adding a finite-range repulsive step function
potential which will push the bound states just above threshold
yielding resonances.  As noted above, the bound states need to be
very deeply bound. Accordingly, to get resonances just above
threshold, one must tune the strength of the repulsive interaction
to very high accuracy to cancel out the binding, leaving behind
barely unbound resonances. However, in principle there is nothing
to prevent one from arranging a system with all of this
fine-tuning done as accurately as one wishes, yielding as many
resonances as one wants as close to threshold as desired.

An example of a two-body central potential that has the desired
properties is
\begin{equation}
\label{eq:Potential} V(r) = -b \sum^{N}_{k=1}{\delta(r-\frac{k
L}{N})} + V_0 \theta(r-(L+\frac{L/N})),
\end{equation}
where $r$ is the distance between fundamental particles, $L$ is
the range of the potential, $b$ is the strength of each of the $N$
delta functions, and the delta functions are raised on a potential
step of height $V_0$. The additional factor of $L/N$ in the step
potential is intended to extended the range of the potential just
beyond the last delta function.  This ensures that potential is
identical in the neighborhood of each delta function. The $\delta$
functions in the potentials should be thought of as very deep,
narrow potential wells---where the details of how this is done
becomes irrelevant provided the width is much smaller than all
other scales in the problem. One can imagine tuning the parameters
in the interaction of Eq.~(\ref{eq:Potential}) (that is, choosing
$b$ and $V_0$) so that any ``would-be" bound states become barely
unbound, turning into low-energy, long-lived resonances. In
Appendix A, we give some numerical evidence that it is possible to
tune the parameters of the two-body interaction of
Eq.~(\ref{eq:Potential}) to create an arbitrary number of nearly
degenerate low-energy resonances.

Qualitatively, one expects that the many different resonant states
will behave as if they were the different species in a
multi-species fluid. However, since these states are resonant
states and not bound states, they eventually decay back into the
fundamental particles, meaning this really is an interacting
single-species gas rather than a multi-species gas. Furthermore,
since the fundamental particles are absolutely stable, this system
describes a stable fluid.

For the system to be of a single species, it is critical that the
system does not have any three- or higher-body bound states.
Given the singular nature of delta functions, one might worry
that the Hamiltonian for three-body or higher-body Hilbert spaces
might be unbounded from below, yielding arbitrarily deeply bound
states. By regulating the delta functions and treating them as
finite width wells, it should become readily apparent that this
will not occur in the zero width limit with fixed resonance
positions. Yet, it is not immediately apparent whether or not the
system, as given, supports three- or higher-body bound states. To
ensure that such states are excluded from our system we also
impose a three-body repulsive potential. We choose the three-body
interaction $V_{3}(\vec{r_1},\vec{r_2},\vec{r_3})$ to be
\begin{equation}
\begin{split}
V_{3}(\vec{r_1},\vec{r_2},\vec{r_3}) & = V_3 \Theta(R -
\text{max}[l_1, l_2, l_3]), \\
l_1 & = |\vec{r_1} - R_{CM}|, \\
l_2 & = |\vec{r_2} - R_{CM}|, \\
l_3 & = |\vec{r_3} - R_{CM}|, \\
R_{CM} & = \frac{m_1 \vec{r_1} + m_2 \vec{r_2} + m_3
\vec{r_3}}{m_1 + m_2 + m_3}, \label{Eq:3body}\end{split}
\end{equation}
where $V_3$, the strength of the three-body interaction, is a
constant set to be larger than any other energy scale in the
problem, $\vec{r_1}$, $\vec{r_2}$, and $\vec{r_3}$ are the
position vectors of the three interacting particles, $R$ is the
range of the three-body interaction, $R_{CM}$ is the location of
the center of mass, and $l_1$, $l_2$, and $l_3$ are the distances
from the center of mass to the location of each particle. The
range of the three-body interaction range $R$ is chosen to be
larger than the range of the two-body interaction $L$. This
interaction forces the interaction between the fundamental
particles and any resonant state to be that of hard sphere
scattering. Once a two-particle resonance is formed, the
three-body potential above prevents the resonance from being
disturbed by interactions with other particles and prevents the
formation of three-particle resonant states.

\subsection{Constructing a bound on the entropy}

The calculation of the entropy of a strongly coupled many-body
system can be quite difficult. Instead we use a variational
argument which shows that entropy  of the entire system for a gas
of many particles interacting through Eq.~(\ref{eq:Potential}) can
bounded from below. In the next subsection, we will choose a
variational ansatz for which the bound is calculable and show that
the lower bound of the entropy can be made arbitrarily large.

Since the fluid under consideration has a finite temperature, we
can work in the canonical ensemble. Recall that in this ensemble,
with natural units ($k_B=1$), the entropy is given by
\begin{equation}
S = \frac{E}{T} + \log(Z).
\end{equation}
where $E$ is the energy of the system, $T$ is the temperature, and
$Z$ is the partition function. By increasing the step height in
Eq.~(\ref{eq:Potential}), we can tune the system to have only
resonant scattering states, and no two-body bound states.
Similarly by choosing the strength of the repulsive three-body
potential in Eq.~(\ref{Eq:3body}) large enough, we can ensure that
there are no three- or higher-body bound states.  This means that
all of the possible configurations of the fluid must have positive
energy. Therefore, the entropy is bounded by
\begin{equation} \label{eq:entropy}
S \geq \log(Z).
\end{equation}
Just as with the entropy, the partition function is difficult to
calculate directly, but the partition function is also bounded
from below.

Recall that in the canonical ensemble the partition function is
given by
\begin{equation} \label{eq:part}
Z = \text{Tr}\, (\exp[-\beta \hat{H}]),
\end{equation}
where $\hat{H}$ is the Hamiltonian operator for the system and
$\beta$ is the inverse temperature. In order to compute the
partition function, one typically needs to use a complete basis
for the Hilbert space of the system.  Since the Hamiltonian is
Hermitian, the operator $\exp[-\beta \hat{H}]$ is positive
semi-definite. This implies that the partial trace over any
arbitrary subspace of the Hilbert space gives a lower bound on
$Z$, termed $Z_{\text{sub}}$.  Choosing such a subspace amounts to
choosing a variational ansatz for the class of configurations of
the fluid: a calculation of the partition function within the
variational ansatz is equivalent to the partition function of some
subspace of the complete Hilbert space. Furthermore, the relation
of the partition functions holds for the logarithm of the
partition function as well,
\begin{equation} \label{eq:loga}
\log(Z) \geq \log(Z_{\text{sub}}).
\end{equation}
Combining Eqs.~(\ref{eq:entropy}) and (\ref{eq:loga}) yields
\begin{equation} \label{eq:entbound}
S \geq \log(Z_{\text{sub}}).
\end{equation}
This shows that the entropy of the entire system is bounded from
below by $\log(Z_{\text{sub}})$. By working with a  variational
ansantz for which the partition function $Z_{\text{sub}}$ is
calculable, we can compute a lower bound on the entropy of the
fluid.

\subsection{Calculating the partition function}

In this subsection we choose a variational ansatz for the system
for which the calculation of the lower bound for the entropy is
tractable. The particular configuration of the system that we
consider is picked entirely for computation ease and is a highly
unlikely one. This merely ensures that the true entropy may be
well above our computed lower bound.

\begin{figure}[htb]
\includegraphics[scale=.3]{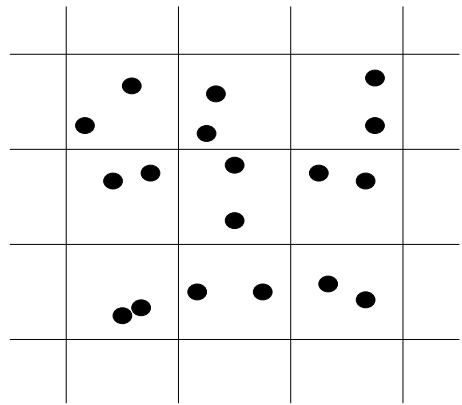}
\caption{\label{fig:cells} As a variational ansatz, we picture the
fluid's volume to be divided into cells with exactly two particles
in each cell.}
\end{figure}

\begin{figure} [htb]
\includegraphics[scale=.2] {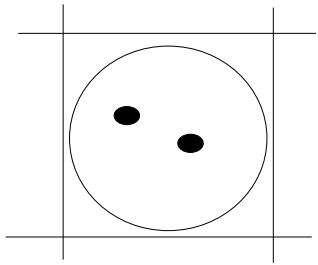}
\caption{\label{fig:cell} A close up view of one particular cell
with the drawn circle representing the constraints on the
particles wave function imposed by the boundary conditions.}
\end{figure}

Consider dividing the volume occupied by the fluid into cells. For
our variational ansatz, we will choose to have exactly two
particles in each cell. The total wave function for this ansatz
can be constructed out of the wave function for each cell as:
\begin{equation}
\Psi_{\text{total}}(\vec{r_1}, \vec{r_2}, \ldots) = \hat{S}
\prod_{\text{cells}  \, i} \Psi_{i}(r_{2i-1},r_{2i})
\end{equation}
where $\Psi_{\text{total}}$, the wave function of the entire
fluid, is a function of the position of every fundamental particle
in the fluid, $\hat{S}$ is an operator which symmetrizes the wave
function under the exchange of any two particles to impose the
exchange symmetry of bosons, and $\Psi_{i}$ is the (two particle)
wave function of each individual cell, and they are summed over
all of the cells. An illustration of the cell decomposition of
the fliud is given in Fig.~\ref{fig:cells}.
To make the computation of the entropy easier, we further restrict
the configurations so that wave function for each cell has the
relative coordinate and center of mass coordinate completely
uncorrelated. With this choice, the wave function for a cell can
be written as
\begin{equation}
\Psi_{\text{cell}}(\vec{r},\vec{R}) = \Psi_{\text{rel}}(\vec{r})
\Psi_{\text{CM}}(\vec{R}),
\end{equation}
where $\vec{r}$ is the relative coordinate, $\vec{R}$ is the
center of mass coordinate, $\Psi_{\text{rel}}$ is the wave
function associated with the relative coordinate, and
$\Psi_{\text{CM}}$ is the wave function associated with the center
of mass. Our ansatz is subject to one further condition: namely,
the following (Dirichlet) boundary conditions:
\begin{equation}\label{eq:bc}
\begin{split}
\left.\Psi_{\text{rel}}(\vec{r})\right|_{r \geq r_{\text{max}}} =
0,
\\
\left.\Psi_{\text{CM}}(\vec{R})\right|_{R \geq R_{\text{max}}} =
0,
\end{split}
\end{equation}
where $r_{\text{max}}$ and $R_{\text{max}}$ are the maximum
relative coordinate and center of mass coordinate, respectively,
that is allowed by a given cell. We take $r_{\text{max}}> L$, so
that the maximum relative coordinate is beyond the range of the
two-body interaction. These boundary conditions ensure that for
this particular ansatz the fundamental particles only interact
within a given cell, and that each cell is isolated from all other
cells. This isolation implies that the two-body interaction plus
the boundary conditions give the dominant contribution to the
partition function within the subspace that we are considering. A
pictorial view of the constraints of the boundary conditions can
be seen in Fig.~\ref{fig:cell}. This highly restrictive ansatz is
certainly an unlikely configuration of the fluid, but it is a
valid variational ansatz; such configurations are present in the
complete Hilbert space.

Having chosen an ansatz for the wave function of the fluid, we can
compute the corresponding partition function. The arguments of the
preceding subsection showed that since the fluid that we consider
has only positive energy states, the entropy of the entire system
will be larger than the logarithm of the partition function
calculated in this ansatz. We have isolated each cell by imposing
boundary conditions, and it is sufficient to calculate the
partition function of only one cell to exhibit the bound. Since
each cell is identical, the total entropy within the ansatz is the
entropy of one cell times the number of cells.  Accordingly the
entropy density of the fluid is bounded by:
\begin{equation}
s \ge \frac{n}{2}  S_{cell}
\end{equation}
where $n$ is the total density (implying that $n/2$ is the
density of cells, and the factor of $\frac{1}{2}$ is due to our choice of two particles per cell).


In order to show that the entropy density of the fluid is
arbitrarily large, we only have to show that the logarithm of the
partition function $\log(Z_{\text{sub}})$ for one particular cell
in the fluid can be made arbitrarily large. To calculate the
partition function, the energies of the states within each cell
are needed. Since the two-body interaction has a finite range, the
relative coordinate wave function within the cell has two
different forms: one within the range of the interaction,
$\Psi_{\text{in}}$, and one beyond the range of the interaction,
$\Psi_{\text{out}}$. The outer wave function is that of a free
state restricted by the boundary conditions, and can be written as
\begin{equation}
\Psi_{\text{out}}(r) = A \sin(k (r_{\text{max}}-r)),
\end{equation}
where $A$ is a normalization factor, $k$ is the momentum of the
state such that $k = \sqrt{2 \mu E}$ with $\mu$ as the reduced
mass, and $E$ is the energy of the state. The momentum, and
thereby the energy, of the quantum states within the cell can be
calculated by matching the logarithmic derivative at the boundary
between the two wave functions.  The matching leads to the
equation
\begin{equation} \label{eq:boundres}
\left.\frac{\Psi'_{\text{in}}(r)}{\Psi_{\text{in}}(r)}\right|_{r=L}
= \left. - k \cot(k (r_{\text{max}}-r))\right|_{r=L}.
\end{equation}
The solutions of these equations give the energies of the states
within each cell.  Relating this condition to the two-body s-wave
scattering phase shifts yields the condition:
\begin{equation}
k r_{\text{max}}= - \delta(k) + n \pi ,
\end{equation}
where $n$ is an arbitrary integer.  Since the phase shifts pass
rapidly through $\pi$ at each resonance, it should be apparent
that there is one low-lying energy state within this ansatz for
every resonance.

The parameters of the two-body interaction can be tuned in such a
manner that all of the resonant states have nearly degenerate,
arbitrarily low energies. If the resonance energies are fine-tuned
to be very small compared to the temperature of the system. their
contribution to the partition function is only slightly suppressed
by a Boltzmann factor and each resonance contributes nearly unity
to the $Z_{\text{sub}}$.  From the resonant contributions it is
easy to see that
\begin{equation} \label{eq:partscale}
\log(Z_{\text{sub}}) > \log(N) - E_H/T
\end{equation}
where $E_H$ is the energy of the highest-lying resonance.  To the
extent that $E_H \gg T$ and $N$ is large, the inequality is almost
saturated; the logarithm of the partition function of the
restricted system thus scales as $\log(N)$. We illustrate that
this scaling can be realized by providing the results of numerical
calculations in Appendix A.

The bound established in the preceding subsection shows that the
system's entropy density, $s$, is larger than
$\log(Z_{\text{sub}})$. By increasing the number of resonant
states while keeping $E_H$ fixed, the lower bound on the entropy
also increases. Since the number of resonant states in the
two-body interaction can become arbitrarily large, so can the
lower bound on the entropy density.

\subsection{Viscosity and Stability}

To complete the argument that the single-species fluid considered
here can violate the class 2 variant of the KSS conjecture, we
need to argue that the shear viscosity $\eta$ does not grow with
the number of two-body resonant states, $N$ (or, more precisely,
grows slower than logarithmically). Furthermore, it is important
to show the resulting  fluid is stable in order to rule out
variants of the conjecture of both classes 2a and 2b.

The shear viscosity is difficult to calculate for virtually any
strongly-interacting system. Fluids for which the Boltzmann
equation is applicable, there are simplifying arguments that allow
one to calculate the shear viscosity \cite{liboff}. However, due
to presence of long-lived resonant states, the fluid described
here does not satisfy the assumptions of the Boltzmann equation.
Therefore, we know of no way to directly calculate the shear
viscosity analytically.

Heuristically, the resonant states in the system described in this
section can be thought of {\it approximately} as bound states. In
Sec.~\ref{sect:class1}, we showed that the shear viscosity of a
system of bound states need not scale uncontrollably with
additional components to the fluid. Therefore, it is difficult to
believe that the shear viscosity for the approximate bound states
would scale vastly differently than that of a dilute
many-component fluid. The actual difference between the shear
viscosity of the two systems should depend on how well the bound
state approximation is valid, which depends on the resonant state
lifetimes. We have constructed the resonant states of the fluid to
have very long lifetimes. As a result, for the purposes of
understanding the shear viscosity, the approximation that the
resonant states can be considered bound states should be quite
accurate. Therefore the shear viscosity of a fluid of long-lived
resonant states should scale similarly to the viscosity of a fluid
of bound states. Moreover, the shear viscosity of a fluid
typically diverges only when it approaches either a
non-interacting ideal gas, or behaves like the cold limit of a
fluid without a defined melting temperature, such as glass. It is
hard to see how a strongly interacting system, such as the one
described in this paper, with a large number of long-lived
resonant states should approach either one of these limits with
the addition of resonant states. Therefore, the shear viscosity
should remain finite as the entropy is made arbitrarily large,
violating the $\eta/s$ bound. While this is not a mathematically
rigorous argument, it is very hard to see how it can fail.

In discussing shear viscosity, we approximated the system as
though it contained bound states. However, at a fundamental level
there are no bound states, and the fluid is still composed of only
one species. If one wanted to compute $\eta/s$ for this system
numerically, for instance, the relevant degrees of freedom to
simulate would be those of the fundamental particles together with
their interactions, and {\it not} of the resonances.  Since these
fundamental particles are absolutely stable, by construction, the
fluid is stable.


\subsection{Summary of results on class 2}

The preceding arguments show that the entropy, and therefore the
entropy density increases with the number of resonant states. We
have argued that although the calculation of the shear viscosity
for the fluid we described is not tractable, there are strong
heuristic reasons to believe that it will not diverge when one
chooses parameters to force the entropy to diverge. To the extent
one accepts these arguments, one must conclude that the ratio
$\eta/s$ can be made arbitrarily small by increasing the number
of resonant states, violating the conjectured bound on $\eta/s$.

The number of resonant states needed to actually violate the bound
could be extremely large, but the two-body interaction that has
been discussed here can be tuned in such a manner as to produce an
arbitrary number of resonant states. That is, there does not
appear to be a limit inherent in the structure of quantum
mechanics on the number of resonant states that can be constructed
within a finite ranged potential.

We note that if a conjecture is false for stable fluids in some
class of theories, it must be false for metastable fluids as well.
As a result, the fluid that we have described in this section
actually provides a counterexample to all theories of class 2,
both for stable and metastable fluids.

\section{Class 3}
\label{sect:class3}

In the preceding two sections, we examined possible variants of
the $\eta/s$ bound for theories of classes $1$ and $2$ and argued
that it is possible to construct systems in those contexts that
violate the bound through a large Gibbs mixing entropy.  As we
noted in Sec.~\ref{sect:classificationTheory}, however, one might
believe that the structure of quantum field theory (and
specifically, the structure of ``sensible" quantum field theories
such as QCD) may rule out counterexamples based on the Gibbs
mixing entropy. That is, the conjecture that $\eta/s \geq 1/4\pi$
may be taken to apply only to systems that can be described by
``sensible" quantum field theories.  This form of the conjecture
would be associated with class $3$ and is similar to the variants
proposed in Refs.~\cite{KSS2,SS1}. In this section, we will review
a counterexample to this class first presented in
Ref.~\cite{cohen}, and give a more detailed discussion of some of
the subtleties in that analysis. To conclude this section, we
discuss a possible objection by Son \cite{sonComment} to the
applicability of this counterexample to the KSS bound of class 3;
we conclude that the issues raised by Ref. \cite{sonComment}
should not limit the applicability of the counterexample.

As we saw in Sec.~\ref{sect:classificationTheory}, a naive
attempt to construct a system of light mesons with a very large
number of different species by increasing the number of flavors
$N_f$ in QCD resulted in the ratio $\eta/s$ scaling as $N_f^2/\log
N_f$, implying that the bound held in the large $N_f$ limit.
Recall that this scaling of $\eta/s$ was due to the fact that to
preserve asymptotic freedom (and thus ``sensibility''), as $N_f$
was increased, the number of colors $N_c$ also had to be increased
proportionally to $N_f$. The bound then held because the cross
section scaled as $1/N_c^2$ in the large $N_c \sim N_f$ limit.
As it turns out, however, this result is not characteristic of
\emph{all} meson gases. In this section we review a
counterexample first discussed in Ref.~\cite{cohen} for theories
of class 3 by considering a \emph{heavy} meson gas.

\subsection{A heavy meson gas}

Consider a gas of heavy mesons.  Each meson is made from a heavy
quark and a light anti-quark.  For the discussion that follows, we
will assume that the gas is only composed from pseudoscalar heavy
mesons, and will justify this assumption below.  We can produce
many heavy meson species by fixing the number of light quark
flavors to some small value with one being adequate, and scaling
the number of heavy quark flavors, $N_f$, to be large: $N_f =
e^{\xi^4}$, where $\xi$ is a dimensionless scaling parameter. As
in Sec.~\ref{sect:class1}, this scaling is chosen to  ensure that
the Gibbs mixing entropy of the heavy meson gas scales as $\xi^4$,
which is what is necessary to drive the ratio $\eta/s$ to zero. As
before, to ensure asymptotic freedom, we must scale the number of
colors, $N_c$, as we scale the number of heavy flavors, hence $N_c
= e^{\xi^4}$. At this point, in the case of the light meson gas,
the meson-meson cross section was seen to scale as $1/N_c^2$, and
the resulting increase in the viscosity prevented a violation of
the bound.  However, the heavy meson cross section does not scale
in the same way, and the same problem does not arise.

Recall that in the example of a nonrelativistic gas discussed in
Sec.~\ref{sect:class1}, it is important to remain in the
low-density, low temperature regime so that the calculation of
both the entropy and the viscosity is tractable.  In this regime,
two-particle scattering is dominant, and the scattering is
described by a Schr\"{o}diger equation for the relative wave
function $\psi$,
\begin{equation} \label{eq:twobody}
(-\nabla^2 + m V) \psi = m E \psi,
\end{equation}
where $m$ is the mass of each of the the interacting particle,
$V$ is the interaction potential, and $E$ is the energy of
scattering associated with the relative motion of the particles.
The critical point is that the cross section depends only on the
combinations  $m V$ and $m E$, but not on $V$ or $E$ separately.

In the nonrelativistic gas case of Sec.~\ref{sect:class1}, $m V$
is scale independent by construction (since neither $m$ nor $V$
scale with $\xi$), while $m E$ scales like the temperature, since
the typical energy of two-body scatterings within a gas is
proportional to the temperature $T$. Therefore, $m E \sim m T
\sim m T_0 \xi^{-2}$, implying that classical two-particle,
low-energy scattering is dominant (assuming that the density is
sufficiently low, as it is with the scaling relations of
Eq.~(\ref{eq:regime})).  This implies that the cross section
becomes scale independent in the large $\xi$ limit. For the pion
gas, by contrast, the interaction potential $V$ scales as
$1/N_c$, and $N_c$ must be large to maintain asymptotic freedom
in the large $N_f$ limit.  The mass of the pions is scale
independent (as is the mass of all light mesons in large $N_c$
limit), and thus in the pion gas $m V$ scales as $1/N_c$, rather
than being scale independent.  Since $N_c$ has the same scaling as
$N_f$, the cross section becomes small in the large $N_f$ limit,
preventing the arguments given in Sec.~\ref{sect:class1} from
applying to the pion gas. As a result, the pion gas is not a
counterexample to the KSS bound. It is now not hard to see how a
heavy meson gas might evade these problems: the mass of the heavy
mesons can be chosen to scale in such a fashion that $m V$
remains scale independent.

However, if the heavy meson mass were to scale with $\xi$ to keep
the combination $m V$ scale independent, the other important
combination, $m E$, would no longer scale as $\xi^{-2}$ as before
unless the scaling of $T$ were to be changed as well.  One might
be concerned that changing the way $T$ scales may cause the system
to no longer be in the regime of low temperature and low density.
However, by fixing the scaling of $m T$ to preserve the scaling of
Eq.~(\ref{eq:twobody}), the low temperature and low density regime
of Eq.~(\ref{eq:regime}) is simultaneously maintained; by a
judicious choice of scaling we can create a nonrelativistic heavy
meson gas equivalent to the nonrelativistic system in
Sec.~\ref{sect:class1}. The necessary scaling relations will be
discussed below.

To understand the necessary scalings, let us begin by examining
the heavy meson interaction. In the heavy meson gas, the
interactions between heavy mesons at long distances is mediated by
the exchange of light mesons. That is, to leading order, the heavy
meson interaction potential is a Yukawa potential,
\begin{equation}
V(r) \sim g^2 \frac{e^{- M_l r}}{r},
\end{equation}
where $g$ is the heavy meson/light meson coupling, and $M_l$ is
the mass of a light meson. The mass of the light mesons is set by
$\Lambda_{\text{QCD}}$, and we can choose $\Lambda_{\text{QCD}}$,
and hence $M_l$, to be independent of $\xi$. By choosing $M_l$ to
be scale independent, we show that the range of the interaction
becomes scale independent as well. The scale dependence of the
potential strength $V(r)$ is then simply given by $g^2$. In the
large $N_c$ limit, we expect that $g \sim 1/N_c^{1/2}$, and thus
$V \sim 1/N_c$. We can now choose the heavy meson mass, $M_h$ to
scale as $N_c$ so that $M_h V$ remains scale independent, as
desired. The mass of the heavy meson is dominated by the mass of
the heavy quark. By scaling the heavy quark mass appropriately,
the heavy meson mass can be fixed to scale as $N_c$. It is easy to
see that choosing the heavy quark mass, $m_h$, to scale as $m_h =
m_{h_0} e^{\xi^4}$, where $m_{h_0}$ is the scale-independent
portion of the heavy quark mass, will result in the correct
scaling of the heavy meson mass, $M_h = M_{h_0} e^{\xi^4}$, where
$M_{h_0}$ is the scale independent portion of the heavy meson
mass. These scaling relations ensure that the relevant quantity
$M_h V$ remains scale-independent. Note that while this simple
argument was given in terms of a meson-exchange picture, valid at
long distance, the scaling arguments hold quite generally.

In the nonrelativistic gas, we saw that $m T$ scales like
$1/\xi^2$. To be consistent with this in the case of the heavy
meson gas, we can set $T \sim 1/(M_h \xi^2) = T_0
e^{-\xi^4}/\xi^2$ with $T_0$ a $\xi$-independent constant. This
scaling relation preserves the scaling of $M_h T \sim \xi^{-2}$ by
construction, and therefore the relations of Eq.~(\ref{eq:regime})
can be satisfied by choosing the density, $n$, to scale as before,
$n \sim n_0 \xi^{-4}$.  Finally, the light quark mass, $m_l$,
should be scale independent, as is the light meson mass. To
summarize, the parameters of the heavy meson gas must scale as
follows:
\begin{eqnarray}
&{}& N_c   =  e^{\xi^4} \;  \;   N_f =  e^{\xi^4} \;  \; \; \; m_h
= {m_h}_0 \, e^{\xi^4} \; \; \;  \;   m_l \sim  {m_l}_0 \nonumber
\\
&{}& \Lambda_{\rm QCD}  =   {\Lambda_{\rm QCD}}_0 \; \; \; \;  \;
\; n = n_0 \xi^{-4} \; \; \;  \; \;  T = T_0
\frac{e^{-\xi^4}}{\xi^2}.
 \label{eq:scale4}
\end{eqnarray}
With these scaling relations, we can simply repeat the argument
given in Sec.~\ref{sect:class1}, and conclude that the
viscosity $\eta$ scales as $\xi^3$, while the entropy density $s$
scales as $\xi^4$. This implies that the ratio $\eta/s$ scales as
$\xi^{-1}$.  By taking $\xi$ to infinity, we can thus drive the
ratio to zero for a system described by a ``sensible" (that is,
asymptotically free) quantum field theory, violating the KSS
conjecture for a system described by a theory of class $3$.

We now justify the assumption that the heavy meson gas is
dominantly composed from spin-0 pseudoscalar mesons, as opposed
to spin-1 vector mesons as $\xi \rightarrow \infty$. Because of
the scaling relations chosen in Eq.~(\ref{eq:scale4}), the heavy
meson gas is clearly in the heavy quark limit. In this limit, the
pseudoscalar, $H$, and vector, $H^*$, heavy mesons are nearly
degenerate. Therefore, one may naively expect both spin states to
be present in the heavy meson gas. However, the two spin states
have a typical mass splitting on the order of
$\frac{\Lambda_{\text{QCD}}^2}{m_h}$ \cite{isgur}. From the
parameter scaling relations in Eq.~(\ref{eq:scale4}), it is not
hard to see that this mass splitting scales like $\sim e^{-\xi^4}
\Lambda_{\text{QCD}}^2/m_{h0}$. We would expect a heavy meson gas
to contain both the pseudoscalar and the vector form of the heavy
mesons, with their populations determined by a Boltzmann
distribution. Let us consider the ratio of the populations of the
vector mesons to the pseudoscalar mesons. From the Boltzmann
distribution, this ratio is given by
\begin{equation}
\frac{N_{\text{vec}}}{N_{\text{pseudo}}} \sim \frac{e^{-\beta
M_{H*}}}{e^{-\beta M_{H}}} = e^{-\beta (M_{H*}-M_H)},
\end{equation}
where $\beta$ is the inverse temperature. Using
Eq.~(\ref{eq:scale4}), we see that
\begin{equation}
\frac{N_{\text{vec}}}{N_{\text{pseudo}}} \sim e^{- \xi^2
\frac{\Lambda_{\text{QCD}}^2}{m_{h0}}},
\end{equation}
which for large values of $\xi$ reduces this ratio to zero.
Therefore, at large $\xi$ the gas is predominantly composed of
pseudoscalar heavy mesons, and we are well justified in neglecting
the heavy vector mesons.

\subsection{Stability}

As we have done with the other counterexamples, the stability of
the fluid needs to be considered. The system that we have
constructed is actually metastable, and thus is a counterexample
to conjectures of class $3$b. Because of the attractive nature of
the potential between heavy mesons, there are several ways by
which the heavy meson gas may decay. However, the decay time
scales are perimetrically large, implying that the gas is
metastable.

In this context, it is natural to consider whether the heavy meson
gas might be susceptible to decay through the formation of
tetraquarks or other multiple meson states. As is well known, as
one approaches the limit of infinitely heavy quark masses (with
light masses held fixed) bound states of two heavy mesons,
tetraquarks, must exist \cite{manohar,hohler2}. The reason for
this is simple: the color Coulomb interaction between the two
heavy quarks allows the formation of a tightly bound diquark to
which the the two light antiquarks then bind.  An alternative
argument is that there is an effective potential between the two
heavy mesons, the long distance part of which is given by pion
exchange which always has an attractive channel when one includes
both vector and pseudoscalars.  It is a general theorem of
elementary quantum mechanics that any potential with an attractive
region always has two-particle bound states in the limit that the
reduced mass becomes large. Since we are considering the limit of
arbitrarily high masses with our scaling rules, one might expect
that bound tetraquarks will exist at large $\xi$ and lead to
metastability. However, this generic theorem that bound
tetraquarks must exist for large enough heavy quark mass assumes a
fixed number of colors $N_c$. The relevant combination is in fact
$m_h N_c^{-1}$ since $\alpha_s \sim N_c$. Thus, the relevant
parameter is $m_{h_0} \sim m_h N_c^{-1}$. If $m_{h_0}$ is small
enough relative to $\Lambda_{\rm QCD}$, the tetraquark will not
bind. Thus, by a judicious choice of parameters one can always
prevent metastability due to the formation of tetraquarks.

It is plausible that there exist values of $m_{h_0}$ small enough
so that stable tetraquarks do not exist but hexaquarks do.
Presumably by fixing $m_{h_0}$ to be smaller still, one can ensure
that stable hexaquarks also do not exist.  More generally, it
should be possible to ensure that any $k$-heavy meson bound states
up to some fixed $k$ are unbound.    This means that to the extent
the system is metastable, it has a very long lifetime: from the
arguments of Sec.~\ref{sect:class1:stability} it is clear that
$\tau_{\text{meta}}/\tau_{\text{fl}} \sim \xi^{5 (k-1)}$, which
for large $\xi$ is very large indeed.

One might hope that by choosing  $m_{h_0}$ to be small enough, we
could eliminate {\it all} possible modes of rearrangement, and
thus obtain a stable fluid, as opposed to merely a {\it very}
long-lived metastable configuration.  However, this is not the
case: the  process in which  $N_c$ heavy mesons rearrange into  a
heavy baryon and a light anti-baryon is always possible. As we
show in Appendix B, it is energetically favorable for the system
to rearrange itself into a heavy baryon and a light anti-baryon.
The binding energy of a baryon made from heavy quarks scales as
$N_c m_h (N_c \alpha_s)^2$ (the binding energy of a light baryon
is order $\Lambda_{\text{QCD}}$ and is negligible), while the
binding energy of a system of $N_c$ heavy mesons scales as $N_c
\Lambda_{\text{QCD}}$. If the heavy quark mass $m_h$ is large
enough, it is energetically favorable for $N_c$ heavy mesons to
rearrange themselves to form a heavy baryon and a light
anti-baryon.  As a result, the heavy meson gas cannot be
absolutely stable.

However, the time scale for the decay of the gas through this
process is parametrically extraordinarily long.  Recall that the
scaling relations for the heavy meson gas are chosen to keep it in
a low-density regime, and that $N_c$-body interactions (i.e.,
collisions) are necessary to convert $N_c$ heavy mesons into two
baryons.  Such interactions are very rare, and the frequency of
such $N_c$-particle collisions decreases with the density of the
gas. By the standard arguments used previously, the ratio of time
scales if the metastability is due to this mechanism is
astoundingly large, scaling as
\[
\tau_{\text{meta}}/\tau_{\text{fl}} \sim \xi^{5(\exp[\xi^4]-1)}.
\]
For quite modest values of $\xi$ this is an exceptionally long
time.  It is not totally clear that this scaling is relevant
since it may be that an instability due to clumping of some
finite but large number of heavy mesons, $k$, might always occur
before this process sets in.  In any case, the lifetime of the
metastable state can be shown to be extraordinarily large.

There is one more decay mechanism for the heavy meson gas which
should be mentioned. Recall that, in
Sec.~\ref{sect:classificationMeta}, we mentioned two general types
of metastable decays. For most of this paper, we have been
discussing the type where the fluid is locally unstable,
but the time scale of the decay is long. However, the heavy meson
gas may also be metastable in the sense of being locally stable
but globally unstable. In the types of fluids with this sort of
instability, the system will typically remain in the metastable
state for extremely long periods of time, usually due to some
potential barrier, until a large perturbation forces the system
into the lower-energy stable configuration. It may be possible
that the heavy meson gas is an example of this type of fluid, but
since we are not violently perturbing the system externally nor is
there an internal mechanism to do so, the time scale associated
with such decays is very large, {\it i.e.} scaling exponentially
with $\xi$. Therefore, this possibility does not alter our
conclusions.

\subsection{The interplay of metastability and the thermodynamic and hydrodynamic limits}

Son \cite{sonComment} has raised an interesting and subtle issue
regarding the interplay of metastability and the thermodynamic and
hydrodynamic limits for the heavy meson gas.  In doing so, he
argues that because of the peculiar nature of this interplay in
the heavy meson system, it is unreasonable to expect the KSS bound
to apply. If one accepts this argument, then the counterexample
given in this section, while valid on its own terms, does not
provide evidence against the validity of the bound for more normal
systems. However, as discussed briefly in
Ref.~\cite{CohenResponse}, the issue raised in
Ref.~\cite{sonComment} does not appear to remove the heavy meson
system discussed above from the class of theories for which a
sensible bound ought to apply.  Thus, we believe that the
conclusions drawn from the existence of this counterexample do not
need to be altered due to the arguments raised in
Ref.~\cite{sonComment}. In this subsection, we outline the issue
raised in Ref.~\cite{sonComment} and discuss its resolution.

The entropy density of a gas becomes well-defined in the
thermodynamic limit. This means that the size of a system is large
enough to contain a sufficiently large number of particles of each
of the possible particle species in the gas so that the entropy
density becomes well-defined. Let us define $V_{\rm t}$ as the
volume in which (on average) we have one meson of every species:
\begin{equation}
V_{\rm t} \equiv L_{\rm t}^3 \equiv \frac{N_f}{n} = \frac{\xi^4
e^{\xi^4}}{n_0}\;, \label{vt}
\end{equation}
where the final form imposes the scaling rules from
Eq.~(\ref{eq:scale4}). As defined above, $V_{\rm t}$ defines the
characteristic volume scale that is associated with the
thermodynamic limit.  $L_{\rm t}^3$ is the characteristic
`thermodynamic length scale' introduced in Ref.~\cite{sonComment}.
It should be clear that to be in the thermodynamic limit, the
physical volume of the system must be much larger than this
characteristic volume. Clearly, from the $\xi$ scaling in
Eq.~(\ref{vt}), {\it very} large systems are required to achieve
the thermodynamic limit for $s$. We note in passing that other
thermodynamic observables, such as energy density or pressure,
approach their thermodynamic limit much more rapidly than the
entropy density since the themodynamic limit of the entropy
density alone depends on every species being present in a large
numbers; hence, other thermodynamic observables do not require
exponentially large systems. In this respect, the heavy meson
system studied in this section is very unusual.

The viscosity, on the other hand, becomes well-defined in the
hydrodynamic limit. This requires that viscosity measurements be
performed on a length scale $L_h$ or larger, where $L_h$ sets a
lower bound on the scale for which fluid behavior is evident. For
dilute systems, such as the one under consideration here, $L_h$ is
effectively the mean free path, $l_{\text{mf}}$.  Using the
scaling relations in Eq.~(\ref{eq:scale4}) one sees that $L_h \sim
\xi^{4}$.

For common fluids such as water or nitroglycerin, the hydrodynamic
length scale $L_h$ is generally  comparable to, or larger than,
the thermodynamic length scale $L_t$.  For typical dilute gases
with one or a few species of particle, $L_h \ge L_t$, since the
mean free path is much larger than the average interparticle
spacing. The heavy meson gas considered here is quite unusual in
that $L_{\rm t} \gg L_{\rm h}$. Because of this fact and the
metastable nature of the fluid, one might think that the bound
should not apply to such systems, as argued in
Ref.~\cite{sonComment}.

To see the issue, suppose we want to {\it measure} $\eta/s$ for
some system composed of the heavy meson gas.  At first glance
there is nothing associated with the metastable nature of the
fluid to prevent one from doing this to very high accuracy (at
sufficiently large $\xi$).  In order to approach the hydrodynamic
limit for which $\eta$ is well defined, one needs to measure
$\eta$ in a system (or a part of a system) which is large compared
to hydrodynamic length scales.  Since as shown above, the ratio of
the life-time of the fluid to the mean collision time is a
positive power law in $\xi$ (or higher), one can measure the
viscosity over a system much larger than $L_{\rm h}$ long before
the system decays.  Thus, $\eta$ is essentially well-defined as a
hydrodynamic quantity. Similarly, $s$ is essentially well-defined
thermodynamically. The issue of concern here is whether the fact
that $\eta$ is essentially well-defined hydrodynamically is
sufficient for the KSS bound to apply.

One natural perspective is that it ought to be sufficient.  If the
bound is general, one might think that it ought to apply to any
system in which $\eta$ (a hydrodynamical quantity) is essentially
well-defined hydrodynamically, and $s$ (a thermodynamical
quantity) is essentially well-defined thermodynamically. There is
an alternative perspective \cite{sonComment}, however. Since the
bound relates $s$ to $\eta$, it is not unnatural to suggest that
it should only apply when $\eta$ and $s$ are both {\it
simultaneously} well defined in the sense of being simultaneously
measurable in the same system.

If one adopts the latter view, there is a potentially serious
problem.  While the fluid clearly lives long enough to measure
$\eta$ accurately over a hydrodynamic length scale, it is very
likely that the system would decay before $\eta$ could be measured
over the exponentially larger thermodynamic length scale.
Accordingly, Ref.~\cite{sonComment} argues that because $\eta$ and
$s$ cannot be determined simultaneously in the heavy meson system,
the system is not in the class of systems for which the bound is
expected to apply. If this is true, then despite the fact that the
heavy meson gas on its own can violate the inequality $\eta/s \ge
1/4\pi$, it does not undermine the possibility of the existence of
a bound which applies to more `normal' systems arising from
underlying UV-complete field theories even if they are metastable.

{\it A priori}, it is difficult to assess which of the two
perspectives is likely to be correct.  The bound is conjectured
rather than derived, and accordingly its underlying assumptions
are unclear.  Thus, one might worry that if the second perspective
turns out to be correct, and that the bound only applies when
$\eta$ and $s$ are both {\it simultaneously} well defined for the
same system, then the heavy meson example would not serve as a
counterexample.  However, as we will show below, despite the
argument of Ref.~\cite{sonComment} outlined above, this
perspective does {\it not} invalidate the heavy meson counter
example.

The key point is that while the argument was formulated in terms
of lengths scales, the thermodynamic limit depends on {\it
volumes}. Recall that in general, for a system to be in the
thermodynamic limit, the volume of the system is required to be
large enough so that repeated measurements of thermodynamic
quantities produce the same results, and that intensive quantities
should be independent of the volume and the {\it shape} of the
system.  In practice, for dilute systems this means that a system
is effectively in the thermodynamic limit if {\it i.}) the system
is large enough to contain a sufficiently large number of
particles of each species, and {\it ii.}) the system is
characteristically thicker than the thermal wavelength $1/\sqrt{m
T}$ for all particles and in all directions.  The second condition
basically says that quantum uncertainties in where particles are
located are small compared to the size of the system.  This
condition on the thickness of the system is the {\it only}
condition on the length scales of the system, as opposed to the
volume. It is easy to see that for systems with the scaling laws
in Eq.~(\ref{eq:scale4}) the thermal wavelength scales as $\xi^1$.
Since the mean free path is always larger than $1/\sqrt{m T}$ for
the heavy meson gas, condition {\it ii.}) is automatically
satisfied for any system with a thickness of the order of the
hydrodynamic length scale or larger, and the question of whether
the thermodynamic limit is reached depends only on the volume of
the system, and not on the shape.

Given this, the notion of a thermodynamic length scale is not
really well-defined: it depends on the arbitrary choice of a
particular shape for thermodynamic system.  With this in mind,
consider as a simple illustration a non-relativistic,
single-species ideal gas of particles of mass $m$ in equilibrium
at density $n$ and temperature $T$, contained in a rectangular
box. Suppose furthermore that the box is highly asymmetrical---the
dimensions are $W \times W \times t$ with $W \gg t$---and that
condition {\it ii.}) is satisfied: $t \gg 1/\sqrt{m T}$.  Now, the
condition for the system to approach the thermodynamic limit with
$s$ well-defined amounts to the condition that the volume times
the density is much larger than unity.  This is satisfied provided
that $W^2 n \gg 1/t$.

Two observations are in order here.  The first is simply that the
thickness, $t$, need {\emph not} be larger than $n^{-1/3}$ in
order for the system to be in the thermodynamic limit. Indeed by
making $W$ large enough it is possible to take $t$ to be {\it much
smaller} then the interparticle spacing, and still have a
consistent thermodynamic result.  This can be explicitly verified
by very elementary calculations.  The second is that the result
holds regardless of whether the rectangular region is considered
to be a physical box containing the fluid, or merely as a fidicual
volume in a much larger system.

This simple result for a single component fluid is trivially
generalized for a multi-component fluid such as the heavy meson
gas considered in this section.  For the heavy meson gas we can again
consider a slab geometry $W \times W \times t$, and find that the
system is in the thermodynamic limit so far as entropy density is
concerned provided that
\begin{equation}
W^2 \gg \frac{N_f}{n \, t} = \frac{1}{n_0 \, t} \, \xi^4 e^{\xi^4}
\label{Wcond}\;.
\end{equation}

Recall at this stage that measurements of $\eta$ necessarily have
a preferred direction.  One considers a fluid with a velocity
gradient transverse to the direction of fluid flow; $\eta$ is the
ratio of the stress to the magnitude of this gradient.  Suppose
that one wishes to measure the viscosity of the heavy meson fluid
in the slab considered above, and takes the direction of the
gradient to be the short side of the slab ({\it i.e} along the
thickness $t$). The viscosity is essentially well defined
hydrodynamically, provided that {\it a.}) $t$ is much larger than
the typical hydrodynamic scale and {\it b.}) the characteristic
time for momenta to propagate through the thickness $t$ is much
shorter then the decay time of the fluid. Repeated measurements of
$\eta$ will then yield the same result up to very small
fluctuations, so that $\eta$ is well-defined. Equation
(\ref{Wcond}) implies that if we choose $W$ large enough, we can
always ensure that the system is {\it simultaneously} in the
thermodynamic limit with essentially well-defined $s$.  To ensure
that this is true, it is sufficient to  take $W=a \xi^2
\exp(\xi^4/2)$ and $t=b \xi^4$, with $a$ and $b$ sufficiently
large constants.  With this construction, at large $\xi$, we have
a system which violates the KSS bound with $\eta$ and $s$
determined simultaneously and each essentially well defined.
Moreover, if the system we are considering is large enough, then
regardless of its shape, one can always find a fiducial volume for
which $\eta$ and $s$ can be measured simultaneously.

The upshot of this is that the requirement that $\eta$ and $s$ be
determined simultaneously in the same system in order for the KSS
bound to apply does {\it not} rule out the heavy meson system as a
counterexample.  At this point, one might object that the slab
geometry considered is not general.  However, this does not
undermine the counterexample.  The bound is supposed to hold
generally for all systems arising from a ``sensible''quantum field
theory, with $\eta$ and $s$ are essentially well-defined and
measurable simultaneously.  A system composed of the heavy meson
gas in this slab-like geometry proves that this is not true.  The
fact that there exist other geometries in which the system decays
before $\eta$ is determined does not alter this.

\subsection{Class 3a}

The previous counterexample does not rule out class 3a (or $3'$
which is a subclass of class 3a) since it involves a metastable
fluid. However, we should note that these variants of the
conjecture have quite limited domains of applicability. Recall
from Sec.~\ref{sect:classificationApplicability}  that variants of
the conjecture of class 3a do not apply to ordinary fluids such as
water since the quantum field theory underlying water, the
standard model, allows nuclear reactions which can alter the
makeup of the fluid, albeit over {\it very} long times.  However,
by hypothesis for class 3a,  metastablity with arbitrarily long
lifetimes is assumed to be qualitatively different from stability.
If this were not the case then our counterexample to class 3b
would also eliminate 3a since the time scales can be made very
long.

Moreover, the fact that class 3a is of such limited applicability
reduces the amount of evidence available to support this variant
of the conjecture.  Recall that one of the strongest pieces of
evidence for the KSS bound was empirical: everyday fluids like
water appear to respect that bound; no known example violating it
exists.  However, this evidence does not apply to conjectures in
class 3a: for the reasons noted above.

Finally we note that although much of the analysis in this paper
concerns the distinction between metastable and stable fluids, it
is quite reasonable to suppose that that this distinction is
unlikely to be important.  We take the view that, while it is
logically possible for there to be a universal lower bound (of
$1/4\pi$) on $\eta/s$ for \emph{only} stable fluids described by
sensible quantum field theories, it is very difficult to see why
such a lower bound should not apply even approximately to
metastable fluids with arbitrarily long lifetimes.

To summarize, in this section we have described a system that
provides a counterexample to the variant of the $\eta/s$ bound of
class $3$b.  The counterexample system is described by a limit of
QCD, a UV-complete quantum field theory, and is metastable with
an arbitrarily long lifetime.  This counterexample does not apply
to variants of the conjecture of class $3$a (and its subclass
$3'$), but this remaining  variant has a very limited regime of
validity, and has relatively little evidence in its support.



\section{Summary and discussion}
\label{sect:summary}

There have been a number of variants of the conjecture on a
universal lower bound for $\eta/s$ \cite{KSS1,KSS2,SS1}.  After
classifying these variants based on their domains of
applicability, we have critically examined several variants of the
conjectures. The broadest conjecture that has been made is of
class $1$, that $\eta/s \geq 1/4\pi$ for all fluids described by
quantum mechanics.  Of all of the forms of the conjecture, this
one has the strongest empirical evidence in its support. However,
there exist counterexamples to variants of class $1$, as we
discussed in Sec.~\ref{sect:class1}.  The counterexample system
constructed there is the prototype of the other counter-examples
discussed in the paper: the ratio $\eta/s$ is driven arbitrarily
close to zero by tuning a system to have a very large entropy
while the shear viscosity is held fixed.

In Sec.~\ref{sect:class2} we discussed a counterexample to
variants of class 2, that the bound holds for nonrelativistic
systems of one species with spin-$0$ or spin-$1/2$.  By choosing a
peculiar interaction potential and tuning its parameters, we
showed that the entropy of a gas can be made arbitrarily large
while arguing that the shear viscosity can remain fixed, violating
the bound. This form of the conjecture appears to have some
limited empirical support, but the existence of a counterexample
to it suggests that the problem with the bound is not limited to
situations with an exponentially large number of species in the
gas --- a contrived but well-defined interaction potential can
produce systems that will violate the bound.

Lastly, in Sec.~\ref{sect:class3}, we showed that the structure of
``sensible" quantum field theories does not appear to forbid the
construction of systems with the very large number of species
necessary to construct the sort of counterexamples that we have
discussed in the preceding sections.  In particular, we exhibited
a counter-example to conjectures of class $3$b, giving an example
of a metastable gas described by an asymptotically free limit of
QCD that can violate the bound.  We note that the subtle issue
raised in ref.~\cite{sonComment} does not appear to alter this.
While class $3$a is not ruled out, it has little evidence in its
support, and applies to a very limited class of theories. To
illustrate the limits on the applicability of variants of the
conjecture of class $3$a, we reproduce Table II from
Sect.~\ref{sect:classificationApplicability} below, with only the
variants of the conjecture that have not been ruled out shown. It
appears that only exotic fluids like the QGP remain as an example
of fluids that might be constrained by this bound.

\begin{table}[htb]
\begin{tabular}{c|c|c|c|c}
Variant & QGP & $He$ &$H_2O$ & $C_3H_5(NO_3)_3$ \\
\hline
3a. & Y & N & N & N \\
$3^{\prime}$. & Y & N & N & N \\
 \hline
\end{tabular}
\caption{Table showing if each remaining conjecture can be applied
(at least approximately) to either the quark gluon plasma (QGP),
liquid Helium ($He$), water ($H_2O$),liquid nitroglycerin
($C_3H_5(NO_3)_3$); Y(es), N(o)} \label{tab:remainingConjecture}
\end{table}

Finally, there is one class of theories which may respect the
conjectured bound which we have not discussed thus far in this
paper. Since the original conjecture was based on the AdS/CFT
correspondence, the bound may only hold for field theories with
gravity duals in five dimensions (for instance, conformal field theories). As
mentioned in Sec.~\ref{sect:evidence}, there is strong theoretical
evidence that the bound is valid for this class of theories.
Generally speaking, conformal field theories would be included in
class 3, as they are UV-complete ({\it i.e.},``sensible'') quantum
field theories. Though we have demonstrated that the bound need
not be respected for all theories of class 3, it is certainly
possible that it holds for some subclass of UV-complete theories.
Hence, one may argue that field theories with gravity duals are the
``sensible" theories needed to maintain the bound.

We should recall at this stage that the restriction of the bound
(if it is universal) to systems which can be described by
UV-complete relativistic field theories is difficult to justify
from first principles.   After all, the speed of light $c$ does
not appear in the bound, as one might expect if it the result is
coming from the relativistic nature of the underlying field theory. Moreover, from a dynamical point of view, while it is
clear that physics at the UV scale of the field theory might
somehow affect low-energy observables such as  $\eta/s$, it is
very hard to understand how the KSS bound could naturally emerge.

To conclude, it appears that there are counter-examples to the
forms of the conjecture which initially appear to be supported
best; the  remaining forms of the conjecture that there is a
universal bound on $\eta/s$ for some well-defined broad class of
systems outside the original domain of conformal field theories
have both limited applicability and little evidence in their
support.   If the bound is correct despite the apparent existence
of the counter-examples described in this paper, it would have to
be due to some physics beyond the frameworks of quantum mechanics
and quantum field theory.  It is conceivable that the bound has a
justification related to quantum gravity \cite{bekenstein} or
string theory, but given our present level of understanding, it is
very difficult to see exactly how this might come about.

{\it Acknowledgments.}  A.C.,\ T.D.C.,\ and P.M.H.\ were supported
by the D.O.E.\ through grant DE-FGO2-93ER-40762.

\appendix

\section{Numerical results from single species fluid model}

In this appendix, we present some numerical results in support of
the argument that the partition function increases with the number
of resonant states $N$. We neglect the effects of the states
associated with the center of mass motion on the partition
function and the effects of states associated with confining the
wave functions in each cell, as these states are independent of
$N$. The parameters of the potential Eq.~(\ref{eq:Potential}) were
chosen as follows (in arbitrary units):
\begin{equation}
r_{\text{max}} = 1.0001; \quad L = 1; \quad m = \frac{1}{2}; \quad
\hbar = k_{B} = 1.
\end{equation}

Discussed in Sec.~\ref{sect:class2}, as $N$ is increased, one has
to tune $b$ and $V_0$ to produce narrow resonances. In Table IV,
we show the partition function and its logarithm calculated with
increasing $N$ and suitably tuned values of $b$ and $V_0$, at
fixed temperature. The values for $b$ were chosen to ensure the
resonant states were nearly degenerate, {\it i.e.} the larger $b$,
the smaller the spread in energy of the resonant states. The
values of $V_0$ were chosen such that all states were barely
resonant states and not bound states, while the temperature, $T$,
was chosen large compared with the highest resonant state energy
but smaller than the lowest-lying state associated with the
artificial confinement to within a cell.

\begin{table}[htb]
\label{numRes}
\begin{tabular}{|c|c|c|c|c|c|c|}
\hline
$N$ & $b$ & $V_0$ & $T$ & $Z_{\text{sub}}$ & $\ln(Z_{\text{sub}})$ \\
\hline
5 & $100$ & $2,500.5$ & 1600 & $5.06$ & $1.62$ \\
10 & $200$ & $10,001.8$ & 1600 & $10.46$ & $2.35$ \\
15 & $350$ & $30,626.2$ & 1600 & $15.46$ & $2.74$ \\
20 & $480$ & $57,601.9$ & 1600 & $20.98$ & $3.04$ \\
30 & $700$ & $122,505$ & 1600 & $30.09$ & $3.40$ \\
\hline
\end{tabular}
\caption{Numerical results showing the increase in the partition
function $Z_{\text{sub}}$ calculated using the variational ansatz.
$N$ is the number of resonant states; $b$ is the strength of each
delta function well in two-body interaction; $V_0$ is the strength
of energy plateau that creates resonant states in the delta
function wells; $T$ is the chosen temperature.} \label{results}
\end{table}

Note that the partition function and its logarithm scales with
larger number of resonant states as expected by
Eq.~(\ref{eq:partscale}). To further illustrate this, we plot the
partition functions and their logarithms in Figs.~3 and 4, along
with linear and logarithmic best-fit curves, respectively.
\begin{figure}[htb]
\includegraphics[scale=.8]{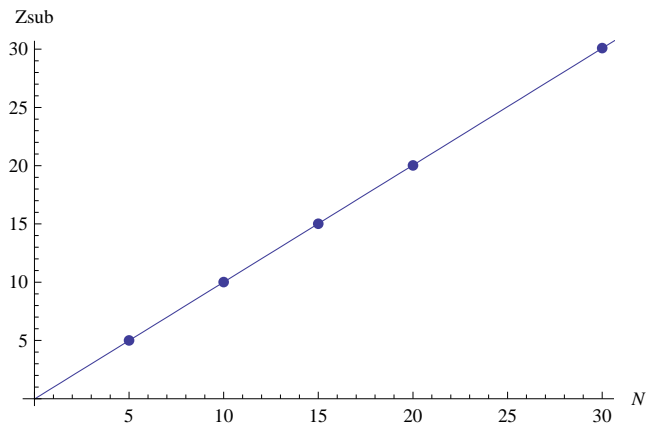}
\caption{Graph of the calculated partition
function and a linear best-fit to the data.}
\label{fig:part}
\end{figure}
\begin{figure}[htb]
\includegraphics[scale=.8]{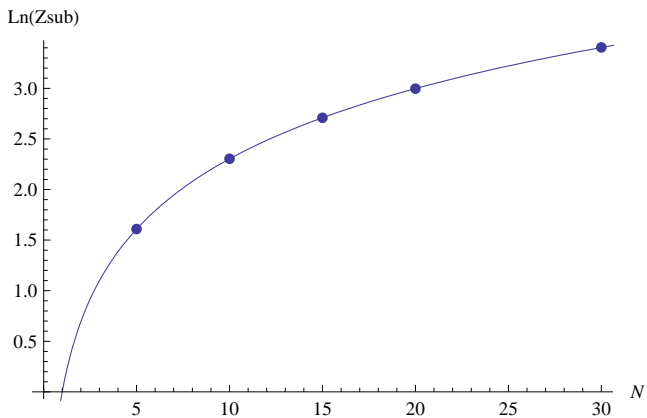}
\caption{Graph of calculated logarithm of the parition function
and a logarithmic best-fit to the data.}
\label{fig:partLog}
\end{figure}

This numerical data supports  the argument that by increasing the
number of resonances in the potential of Eq.~(\ref{eq:Potential}),
it is possible to increase the lower bound on the partition
function of the system, and thereby increase the lower bound on
the entropy.

\section{Metastability of heavy meson gas}

As argued in Sec.~\ref{sect:class3}, the heavy meson gas that we
discussed there is not susceptible to decay via tetraquark and
higher-quark state formation.  However, as we show in this
appendix, one cannot tune the parameters of the heavy meson gas to
prevent the formation of all bound states. It turns out to be
energetically favorable for $N_c$ heavy mesons to rearrange their
quark content to form a heavy baryon and a light anti-baryon.

In order to show that the baryon configuration is more stable than
the meson configuration, we must demonstrate that the binding
energy is larger for the baryons  than for the mesons. The binding
energy of a single heavy meson should be $\sim
\Lambda_{\text{QCD}}$. However, a rearrangement into baryons needs
$N_c$ heavy mesons because the baryons must be color singlets;
thus the total binding energy of the heavy mesons is $N_c
\Lambda_{\text{QCD}}$.  The binding energy of a light anti-baryon
must be $\sim \Lambda_{\text{QCD}}$, as $\Lambda_{\text{QCD}}$ is
the only scale for the light quarks.


Next, we must determine the binding energy of the heavy baryon. As
noted long ago by Witten \cite{Wit}, a baryon composed entirely of
heavy quarks in the large $N_c$ limit is described accurately in
the Hartree approximation for nonrelativistic quarks interacting
via a color Coulomb potential. We can calculate the binding energy
relative for the heavy baryon parametically via a viral theorem
for this Hartree system. The single particle Hartree Hamiltonian
$H$ for the system with $N_c$ heavy mesons has a kinetic term and
a potential term:
\begin{equation}
\hat{H} = \hat{T} + \hat{V}. \label{eq:varHam}
\end{equation}
The kinetic term, $\hat{T}$, can be expressed in the typical
manner in terms of the Laplacian,
\begin{equation}
\hat{T} = \frac{-1}{2 m_h} \nabla^2,
\end{equation}
where $m$ is the mass of the heavy quark. The potential term,
$V(r)$, can be written using a mean field approximation. The
leading order one-gluon potential that binds heavy quarks together
has the form of a Coulombic potential, so in the mean field
approximation, the potential term can be written as
\begin{equation}
\hat{V} = N_c \alpha_s \int \frac{\rho(r')}{|r-r'|} d^3 r',
\end{equation}
where $\alpha_s$, the strong coupling constant, has been factored
outside the integral, and $\rho(r')$ is the particle density for
one of the heavy quarks; the external factor is technically
$N_c-1$ and indicates that each of the remaining quarks
contributes. We will denote the exact single particle ground-state
wave function of the Hamiltonian of Eq.~(\ref{eq:varHam}) by
$\Psi(r)$.

In order to parameterize the energy of the ground state, instead
of $\Psi(r)$, we choose $\varphi(\lambda r)$ as a variational
ansatz, with $\lambda$ is the variational parameter.  If we choose
the form of $\varphi(\lambda r)$ such that it happens to reproduce
the form of the exact Hartree solution, the variational equations
with respect to $\lambda$ will yield an exact relation, {\it i.e.}
a viral theorem. The ground state energy can now be determined by
minimizing the Hamiltonian with regards to $\lambda$.   In order
to perform this minimization, we must first determine how the
kinetic and potential terms scale with $\lambda$. Using the change
of variable $R \equiv \lambda r$ it is easy to see that the
kinetic term must scale as
\begin{equation}
T(\lambda) \equiv \langle \varphi(\lambda
r)|\hat{T}|\varphi(\lambda r)\rangle= \lambda^{-2} T(1) \equiv
\frac{T_0}{m_h \lambda^2},
\end{equation}
we have factored out $1/m_h$ so that $T_0$ is independent of
$\lambda$, $m_h$, $\alpha_s$, and $N_c$.

To find the $\lambda$-scaling of the potential energy term, we
first note that the single particle density can be written in
terms of the wave function $\Psi(r)$ as
\begin{equation}
\rho(r) =\Psi^*(r) \Psi(r).
\end{equation}
When we consider the scaling parameter, the density can be written
as,
\begin{equation}
\rho(r) = \lambda^3 \varphi^*(R) \varphi(R) = \lambda^3 \rho(R)
\end{equation}
where $R= \lambda r$ once again and the factor of $\lambda^3$
comes from the normalization of the variational ansatz. With this
expression for $\rho(r)$, the scaling of the potential is given by
\begin{eqnarray}
V(\lambda)& \equiv \langle \varphi(\lambda
r)|\hat{V}|\varphi(\lambda r)\rangle =N_c \alpha_s \int \frac{\rho(R')}{|\lambda r - R'|} d^3 R'  \nonumber \\
& = N_c  \alpha_s \frac{1}{\lambda} \int \frac{\rho(r')}{|r - r'|}
d^3 R'  \equiv \lambda^{-1} N_c \alpha_s V_0,
\end{eqnarray}
where we first did a change of variables from $R'$ to $r' =
\lambda^{-1} R'$, and then factored $\lambda$ out of the integral,
leaving the factor $V_0$ independent of $\lambda$, $m_h$, $N_c$,
and $\alpha_s$. This has the effect of explicitly showing the
$\lambda$ scaling of the potential energy.

Using the above results, the Hamiltonian now takes a form where the $\lambda$ scaling is fully explicit:
\begin{equation}
H(\lambda) = \frac{T_0}{m_h \lambda^2} + \frac{N_c
\alpha_s}{\lambda} V_0.
\end{equation}
It is now easy to minimize this equation with respect to
$\lambda$,  and the variational estimate of the ground state turns
out to be
\begin{equation}
\label{eq:gs} E_0 =  - m_h (N_c \alpha_s)^2 \frac{V_0^2}{4T_0},
\end{equation}
where $-E_0$ is the binding energy for one quark.   $E_0$ is
expected to be negative, indicating a bound state; the binding
energy of the heavy baryon is $BE = - N_c m_h (N_c \alpha_s)^2
V_{0}^{2}/(8 T_0)$. Since by construction both $T_0$ and $V_0$ are
factors depending only on the form of the variational wave
function and independent of $\lambda$, $m_h$, $N_c$, and
$\alpha_s$ one has the following scaling of the binding energy
with the parameters of the problem
\begin{equation}
BE(\text{heavy baryon}) \sim -  N_c m_H (N_c \alpha_s)^2.
\end{equation}
Note that $N_c \alpha_s$, the square of the `t Hooft coupling is
independent of $N_c$.

At this point we observe that before the rearrangement, the heavy
mesons had a binding energy of $N_c \Lambda_{\text{QCD}}$, while
after the rearrangement, the heavy baryon has a binding energy of
$N_c m_h (N_c \alpha_s)^2$, while the light anti-baryon has a
characteristic binding energy of $\Lambda_{\text{QCD}}$, which is
negligible by comparison. Since the `t Hooft coupling constant
scales like $\sim 1/\log^2(m_h/\Lambda_{\text{QCD}})$, the binding
energy for the heavy baryon will always be perimetrically larger
than for the $N_c$ heavy mesons (which also scales as $N_c
\Lambda_{\rm QCD}$) for a large enough value of the heavy quark
mass. Therefore, this rearrangement of quarks is always
energetically favorable, and thus the heavy meson gas is
metastable relative to this rearrangement.


\begin{thebibliography}{99}
\bibitem{chs}
The coefficient, $C_{hs}$, can be calculated in the Chapman Enskog
expansion as outlined in \cite{liboff} and is $5/ (16 \sqrt{\pi})$
to leading order.
\end{thebibliography}
\end{document}